\documentclass[prb, superscriptaddress, twocolumn, 10pt, showpacs]{revtex4}  
\usepackage{amssymb}
\usepackage{amsmath}
\usepackage{bbm}
\usepackage{graphicx}
\usepackage{color}

\begin{document}

\author{Philipp Werner}
\affiliation{Department of Physics, University of Fribourg, 1700 Fribourg, Switzerland}
\author{Martin Eckstein}
\affiliation{Department of Physics, University of Erlangen-N\"urnberg, 91058 Erlangen, Germany}

\title{Nonequilibrium RIXS study of an electron-phonon model}

\begin{abstract}
We use the nonequilibrium dynamical mean field theory formalism to compute the equilibrium and nonequilibrium resonant inelastic X-ray scattering (RIXS) signal of a strongly interacting fermionic lattice model with a coupling of dispersionless phonons to the total charge on a given site. In the atomic limit, this model produces phonon subbands in the spectral function, but not in the RIXS signal. Electron hopping processes however result in phonon-related modifications of the charge excitation peak. We discuss the equilibrium RIXS spectra and the characteristic features of nonequilibrium states induced by photo-doping and by the application of a static electric field. The latter produces features related to Wannier-Stark states, which are dressed with phonon sidebands. Thanks to the effect of field-induced localization, the phonon features can be clearly resolved even in systems with weak electron-phonon coupling.
\end{abstract}

\date{today}


\maketitle

\hyphenation{}

\section{Introduction}

Resonant inelastic X-ray scattering (RIXS) is a photon-in photon-out technique which provides information on the low-energy excitations in solids,\cite{Ament2011rmp} including spin,\cite{Braicovich2009,Schlappa2018} orbital,\cite{Schlappa2012} lattice, \cite{Ament2011, Chaix2017} and charge \cite{Hill1998,Chen2010} excitations. The recent development of time-resolved RIXS\cite{Dean2016,Mitrano2020} thus promises new perspectives on the nonequilibrium states of laser-driven solids, and in particular on the complex interplay between different active degrees of freedom in perturbed strongly correlated materials. To harness the full potential of this experimental technique, theoretical modeling and numerical simulations are important. In electron-phonon systems, for example, one may ask how the interplay between lattice and charge excitations manifests itself in the RIXS spectrum in equilibrium and nonequilibrium situations. 

Such theoretical and numerical calculations pose considerable challenges. While exact diagonalization calculations on small clusters have been successfully used to explain the main features of the equilibrium RIXS spectra for a range of materials,\cite{Tsutsui1999,Chen2010,Schlappa2012,Schlappa2018} their extension to nonequilibrium problems faces the difficulty of representing the properties and evolution of a generic nonequilibrium state on a small cluster. Certain properties of a pump-excited single-band Hubbard model have been successfully studied with this technique,\cite{Wang2020} but the extension of the cluster approach to electron-phonon (or multi-orbital) problems, and to a wide range of nonequilibrium situations is highly challenging. 

An alternative route is to use an embedding scheme, as in dynamical mean field theory (DMFT),\cite{Georges1996} where the solid is described in terms of a correlated site embedded in an electron bath which represents the lattice environment. While momentum resolution is lost in this approach, DMFT-based RIXS spectra for transition metal compounds capture not only the local $d$-$d$ charge excitations, but also fluorescent features associated with a transfer of kinetic energy to other quasi-particles.\cite{Hariki2018,Hariki2020} This DMFT approach can be extended to nonequilibrium set-ups as demonstrated in Refs.~\onlinecite{Eckstein2021} and \onlinecite{Werner2021}. As in standard nonequilibrium DMFT,\cite{Aoki2014} the nonequilibrium state of the lattice is captured by the self-consistently computed nonequilibrium hybridization function of the DMFT impurity problem, with which the RIXS amplitude is evaluated in a post-processing step.\cite{Eckstein2021} Because it only involves the solution of an impurity model, the DMFT approach is numerically cheaper than cluster schemes, and thus can be used to compute the time-resolved RIXS signal of multi-orbital Hubbard models.\cite{Werner2021} Here, we will use a suitably modified implementation of the nonequilibrium DMFT-based RIXS technique to study the RIXS spectra of a Holstein-Hubbard model, both in equilibrium and in nonequilibrium states induced by pulsed or static electric fields. 

The paper is organized as follows: Section~\ref{sec:method} introduces the model and explains the technique used to treat local electron-phonon couplings. The simulation results are presented in Sec.~\ref{sec:results}, where we focus on the RIXS spectra of a Mott insulating system, both in equilibrium (Sec.~\ref{sec:results:eq}), and out of equilibrium in a photo-doped state and under the influence of a strong static electric field (Secs.~\ref{sec:results:noneq} and \ref{sec:results:ws}). Section~\ref{sec:conclusions} contains the conclusions.

\section{Model and method}
\label{sec:method}

\begin{figure}[t]
\begin{center}
\includegraphics[angle=0, width=0.7\columnwidth]{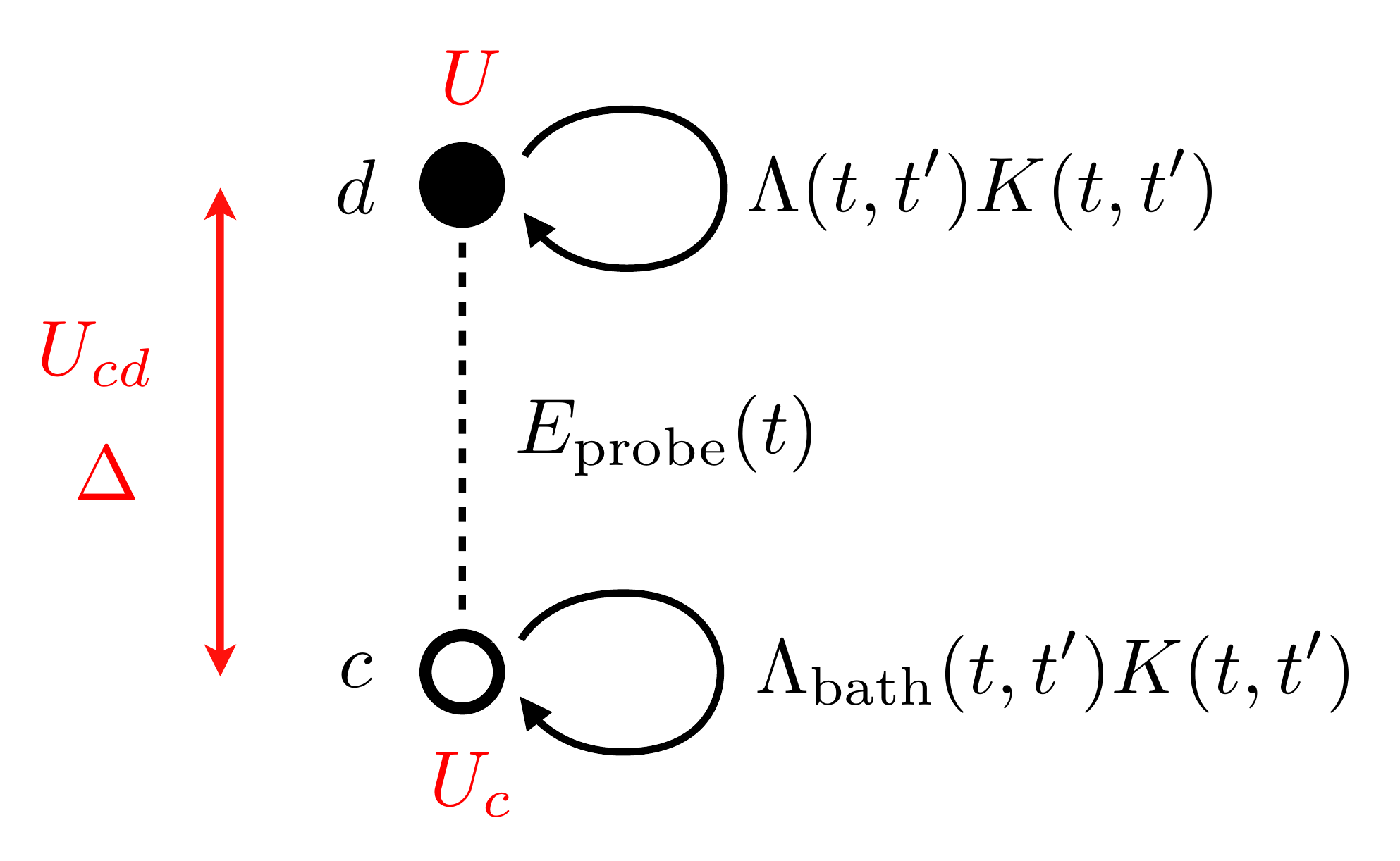}
\caption{Structure of the impurity model for the RIXS calculation. The effective hybridization function of the valence ($d$) orbital, $\Lambda(t,t')K(t,t')$, incorporates the effects of the lattice, the pump pulse, and the phonons. The RIXS probe pulse corresponds to dipolar excitations between the core ($c$) and valence orbital. An equilibrium free-electron bath, corresponding to a hybridization function $\Lambda_\text{bath}$, is coupled to the core level to produce a finite core-hole lifetime. Since this bath changes the number of electrons on the impurity, it is also multiplied with $K(t,t')$.}
\label{fig_impurity}
\end{center}
\end{figure}

We consider a single-band Hubbard model with a local coupling to localized core levels and dispersionless phonons. The Hamiltonian contains the terms
\begin{equation}
H=H_\text{kin}+{\textstyle\sum_i} \big(H_\text{loc,i}^\text{el}+H_\text{loc,i}^\text{el-ph}+H_\text{ph,i}\big),
\label{model}
\end{equation}
where $H_\text{kin}$ describes the hopping of the valence electrons (creation operator $d^\dagger_\sigma$ for spin $\sigma$) between neighboring lattice sites, and the sum is over lattice sites $i$. 
Explicitly, the different terms read
\begin{eqnarray}
H_\text{kin} &=& -v(t) {\textstyle\sum_{\langle i,j\rangle,\sigma}} (d^\dagger_{i,\sigma}d_{i,\sigma}+\text{h.c.}),\\
H_\text{loc}^\text{el} &=& -\mu(n_d+n_c)+\tfrac{\Delta}{2}(n_d-n_c) \nonumber\\
&&+U n_{d,\uparrow} n_{d,\downarrow} + U_c n_{c,\uparrow} n_{c,\downarrow} + U_{cd}n_cn_d \nonumber\\
&& + E_\text{probe}(t){\textstyle\sum_\sigma}(d^\dagger_\sigma c_\sigma+ \text{h.c.})\nonumber\\
&& + H_\text{bath},\\
H_\text{loc}^\text{el-ph} &=& g(n_d+n_c)(b^\dagger+b),\\
H_\text{ph} &=& \omega_0 b^\dagger b,
\end{eqnarray} 
where we used the densities $n_{d,\sigma}=d^\dagger_\sigma d_\sigma$, $n_d=n_{d,\uparrow}+n_{d,\downarrow}$, and similarly for the core electrons (creation operator $c^\dagger_\sigma$, $n_{c,\sigma}=c^\dagger_\sigma c_\sigma$, $n_c=n_{c,\uparrow}+n_{c,\downarrow}$). $v$ is the hopping amplitude between the sites, which becomes time-dependent and complex in the case of an applied electric field.\cite{Turkowski2005} We consider a lattice with a semi-circular $d$-electron density of states of bandwidth $4v(0)$ in the noninteracting equilibrium case.  
$H_\text{loc}^\text{el}$ captures the effects of the chemical potential $\mu$, the core-valence splitting $\Delta$, the Hubbard interactions $U$ and $U_c$, as well as the interaction $U_{cd}$ between the valence electrons and the core electrons (see illustration in Fig.~\ref{fig_impurity}). In addition, we add a dipolar excitation term which allows to describe the excitation of core electrons to the valence orbital by the RIXS pulse (amplitude $E_\text{probe}$, which also incorporates the matrix elements), and  a term $H_\text{bath}$ (free electron bath attached to the core level) which limits the core-hole lifetime. The precise spectral density and coupling of the bath will be defined in terms of the bath hybridization function below. In the spirit of the Holstein model,\cite{Holstein1958} $H_\text{loc}^\text{el-ph}$ describes the interaction of dispersionless phonons (creation operator $b^\dagger$, frequency $\omega_0$, coupling strength $g$) with the total charge on a given lattice site, while $H_\text{ph}$ corresponds to the free phonon Hamiltonian.  

To treat the correlation effects and measure the RIXS signal, we use nonequilibrium DMFT.\cite{Freericks2006,Aoki2014} In this formalism, the lattice model is mapped to a quantum impurity model embedded in a time-dependent bath represented by a hybridization function $\Lambda_{d,\sigma}(t,t')$. This hybridization function determines how electrons hop in and out of the valence ($d$) orbital and mimics the lattice environment. We use the procedure described in Ref.~\onlinecite{Werner2017} to capture the effects of a (pump) electric field, so that the self-consistency relation, which relates the impurity Green's function $G_d$ to the hybridization function, becomes
\begin{align}
\Lambda_{d,\sigma}(t,t') = &v(t)\cos(\phi(t))G_d(t,t')v(t')\cos(\phi(t'))\nonumber\\
&+v(t)\sin(\phi(t))G_d(t,t')v(t')\sin(\phi(t')), 
\end{align}
with $\phi(t)=eaA_\text{pump}(t)$ and $A_\text{pump}(t)$ the vector potential of the electric field $E_\text{pump}(t)=-\partial_t A_\text{pump}(t)$ in a gauge without scalar potential ($e$ is the electric charge and $a$ the lattice constant, which we set to unity).  

The hybridization function $\Lambda_{d,\sigma}(t,t') $ is determined by a standard nonequilibrium DMFT calculation, which may involve the action of a pump electric field on the system, but does not involve any RIXS pulse ($E_\text{probe}=0$). Keeping this hybridization function fixed, the RIXS signal can then be calculated in a second step as described in Ref.~\onlinecite{Eckstein2021}. 

\begin{figure*}[ht]
\begin{center}
\includegraphics[angle=0, width=0.49\columnwidth]{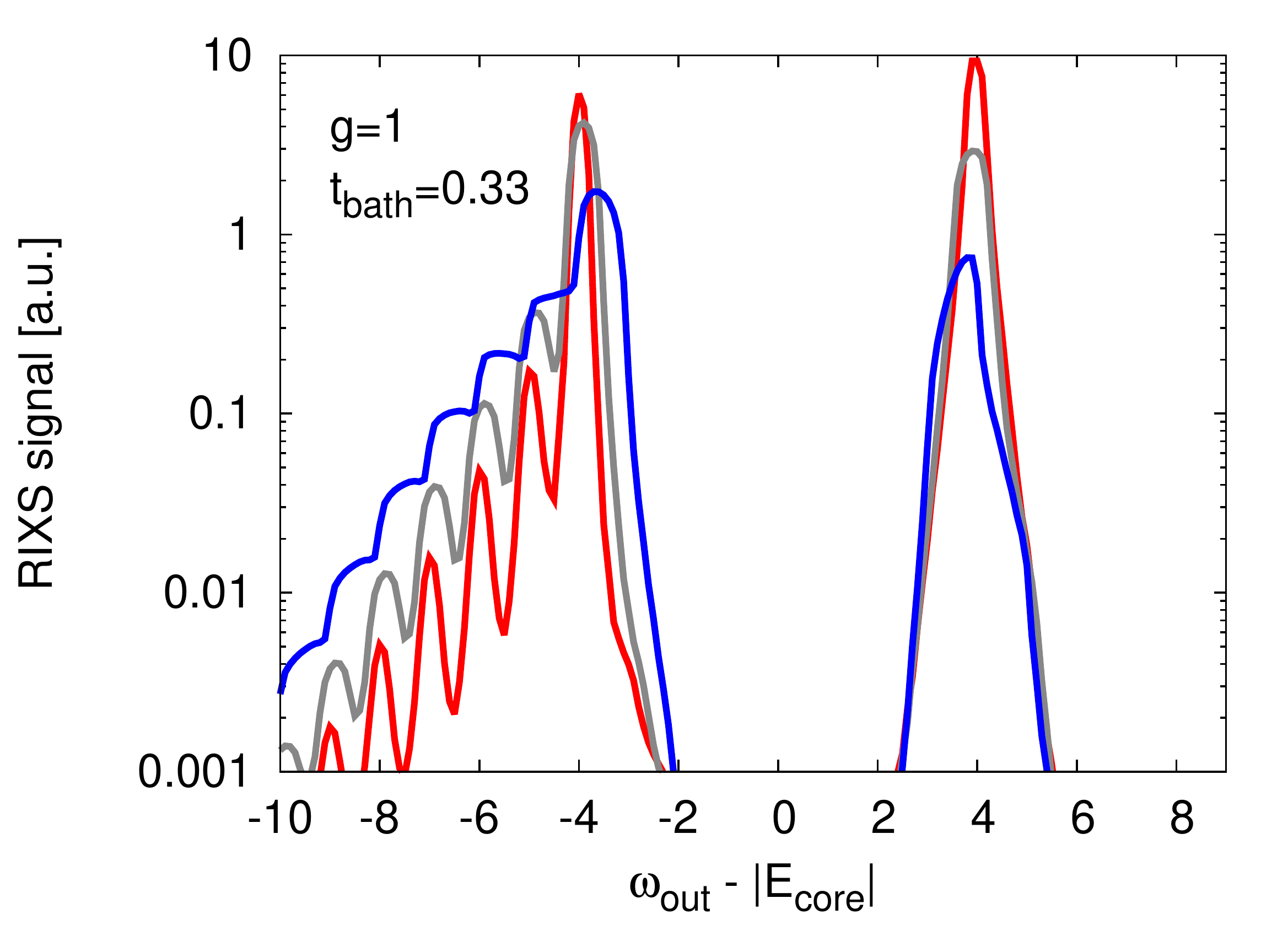}\hfill 
\includegraphics[angle=0, width=0.49\columnwidth]{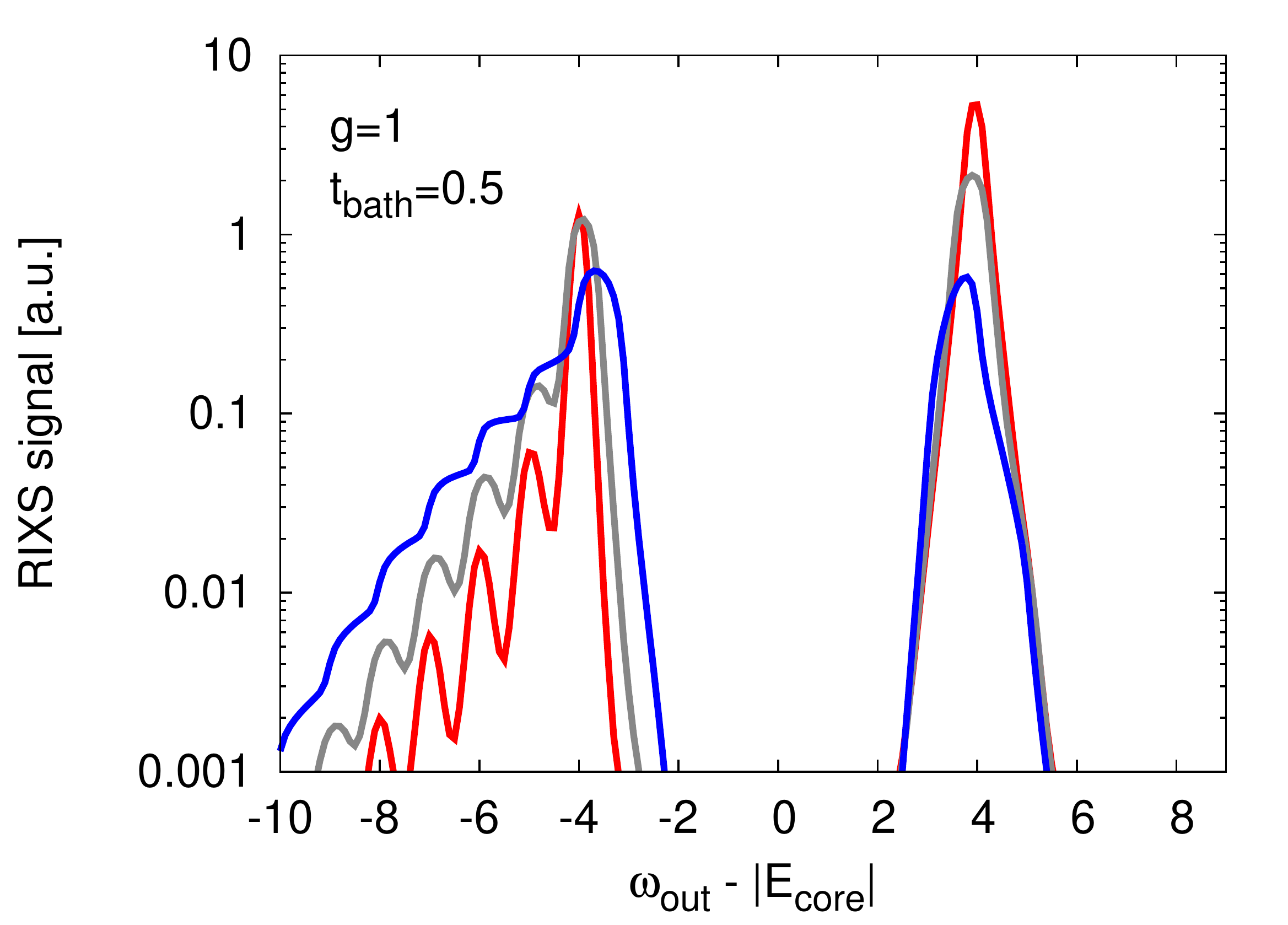}\hfill 
\includegraphics[angle=0, width=0.49\columnwidth]{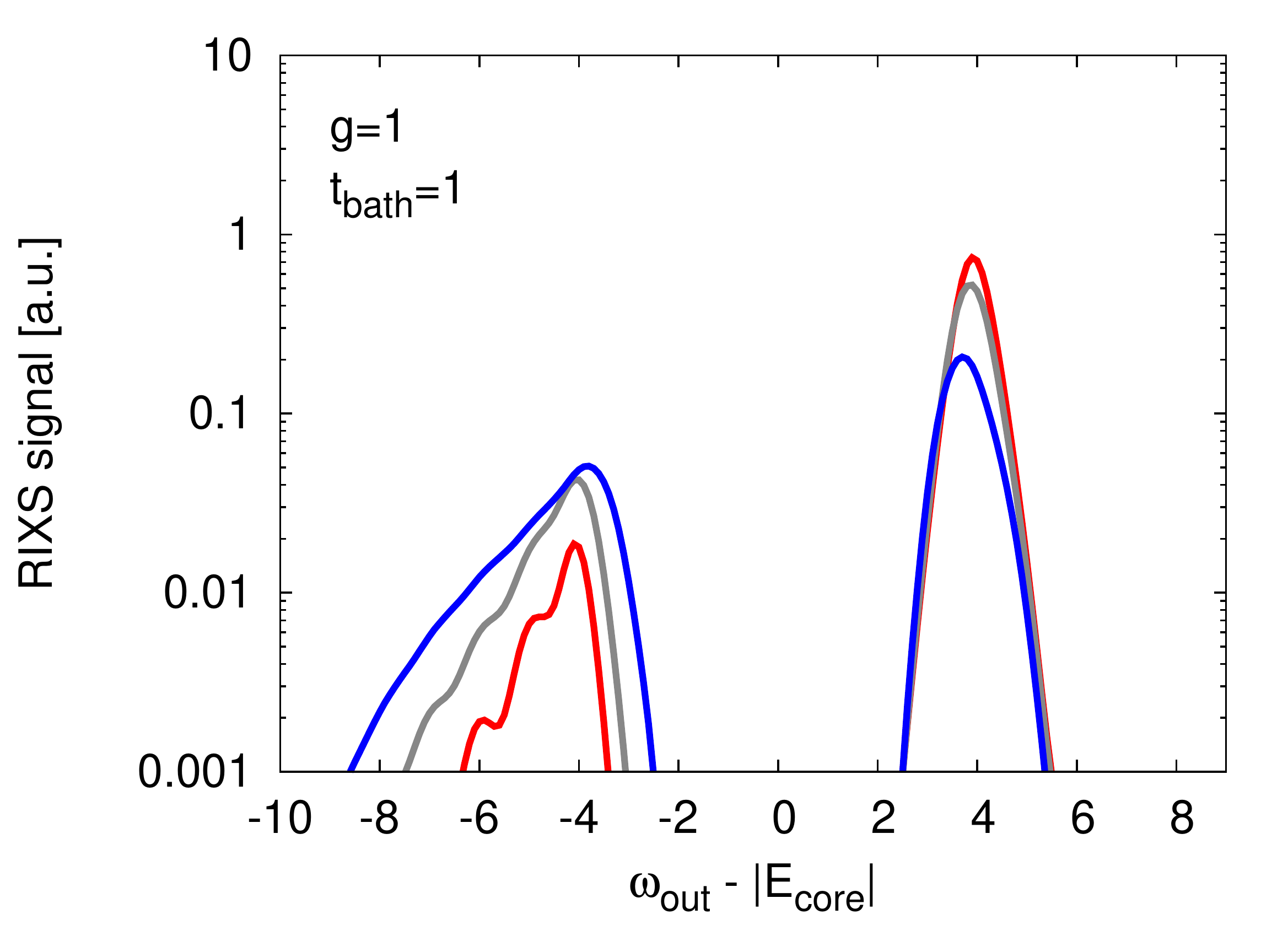}\hfill 
\includegraphics[angle=0, width=0.49\columnwidth]{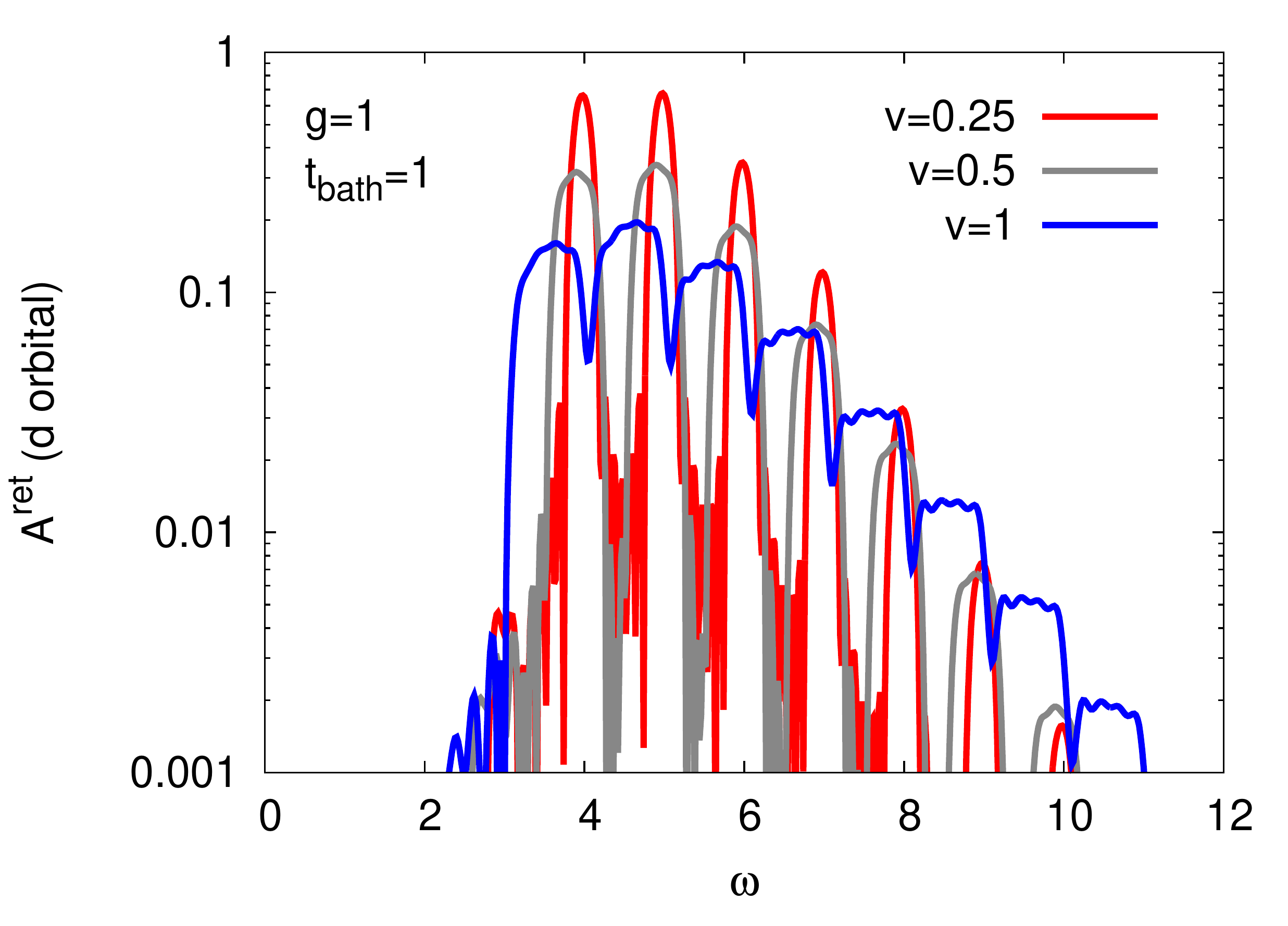}\\ 
\includegraphics[angle=0, width=0.49\columnwidth]{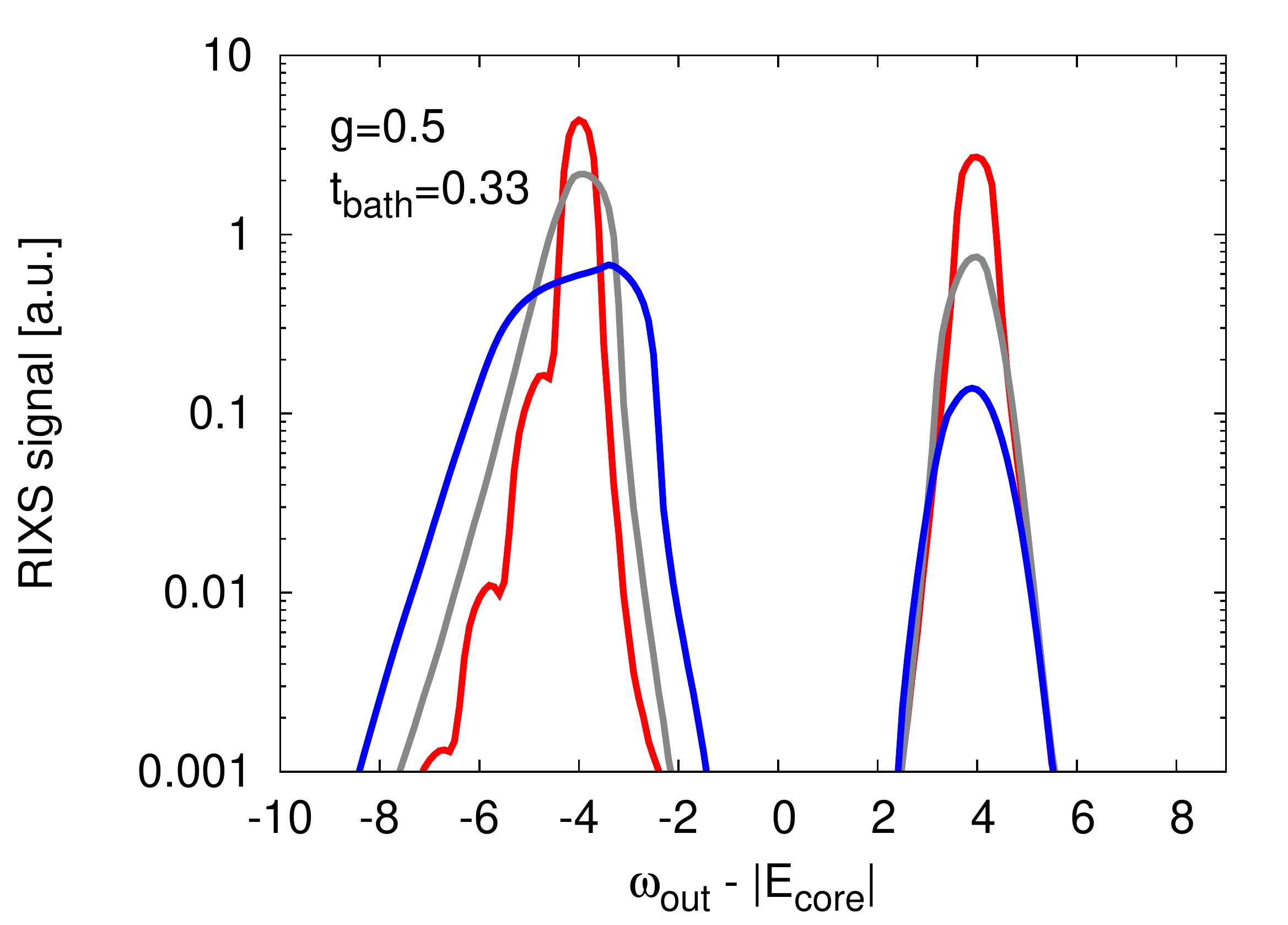}\hfill 
\includegraphics[angle=0, width=0.49\columnwidth]{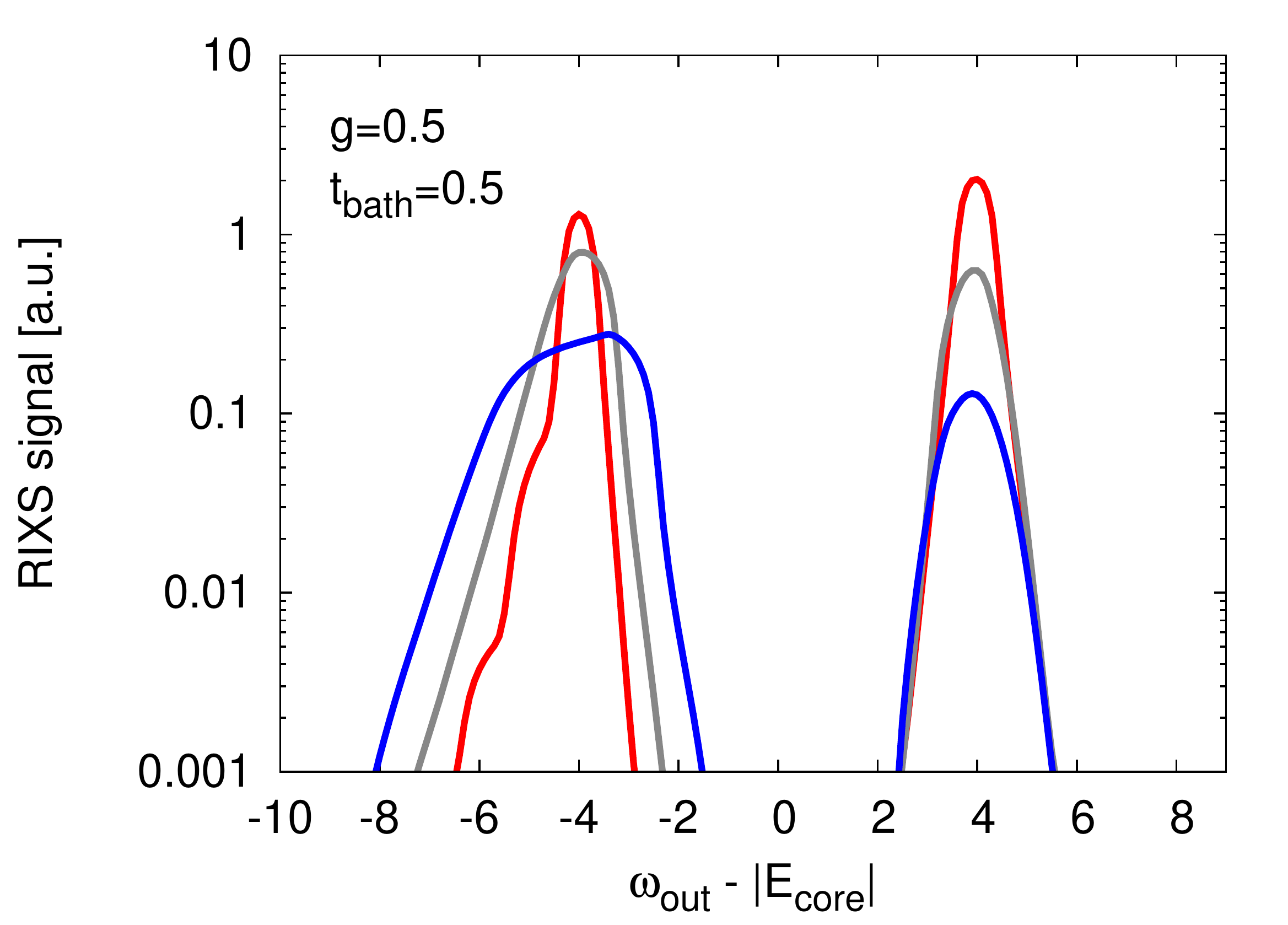}\hfill 
\includegraphics[angle=0, width=0.49\columnwidth]{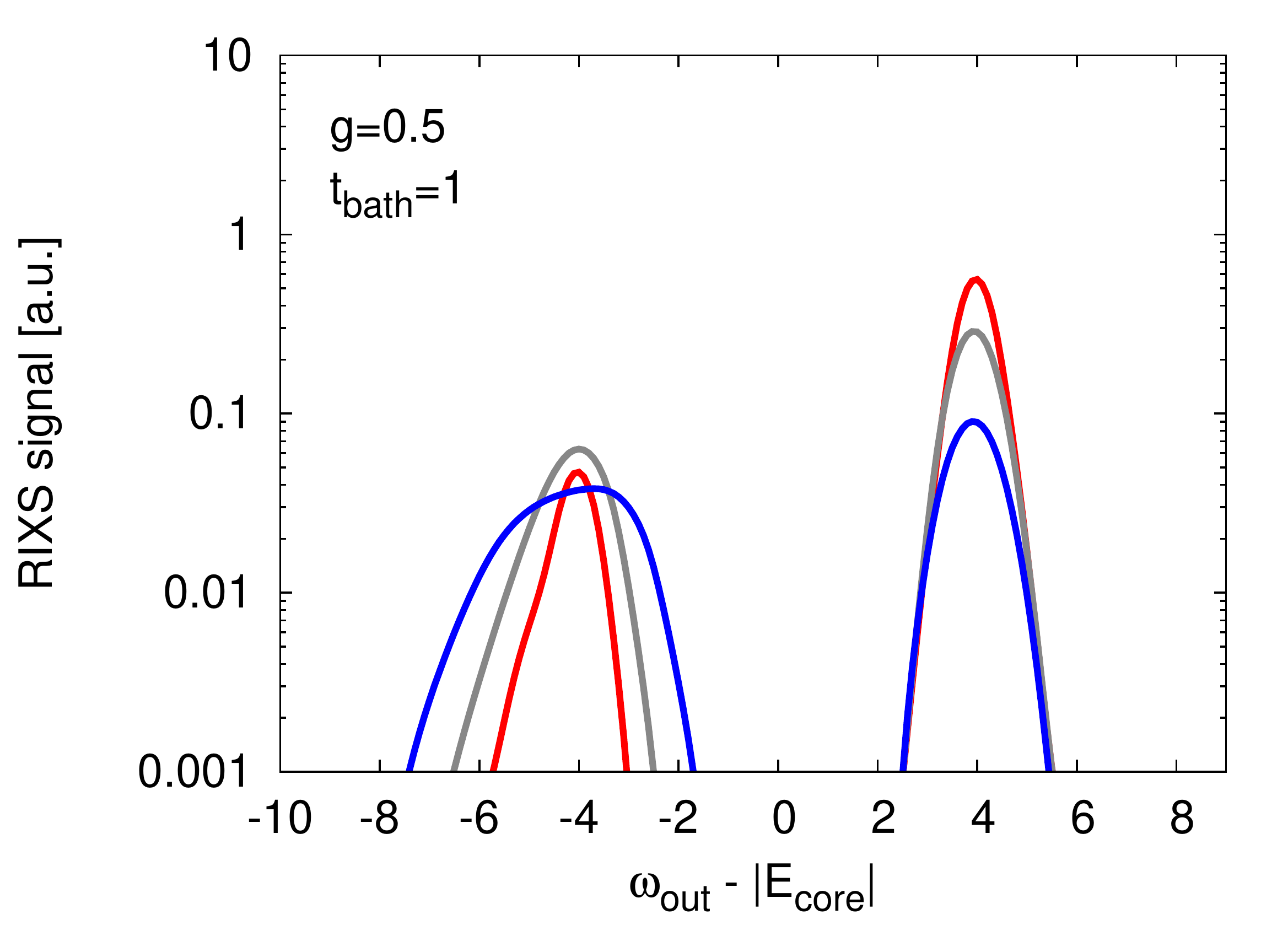}\hfill 
\includegraphics[angle=0, width=0.49\columnwidth]{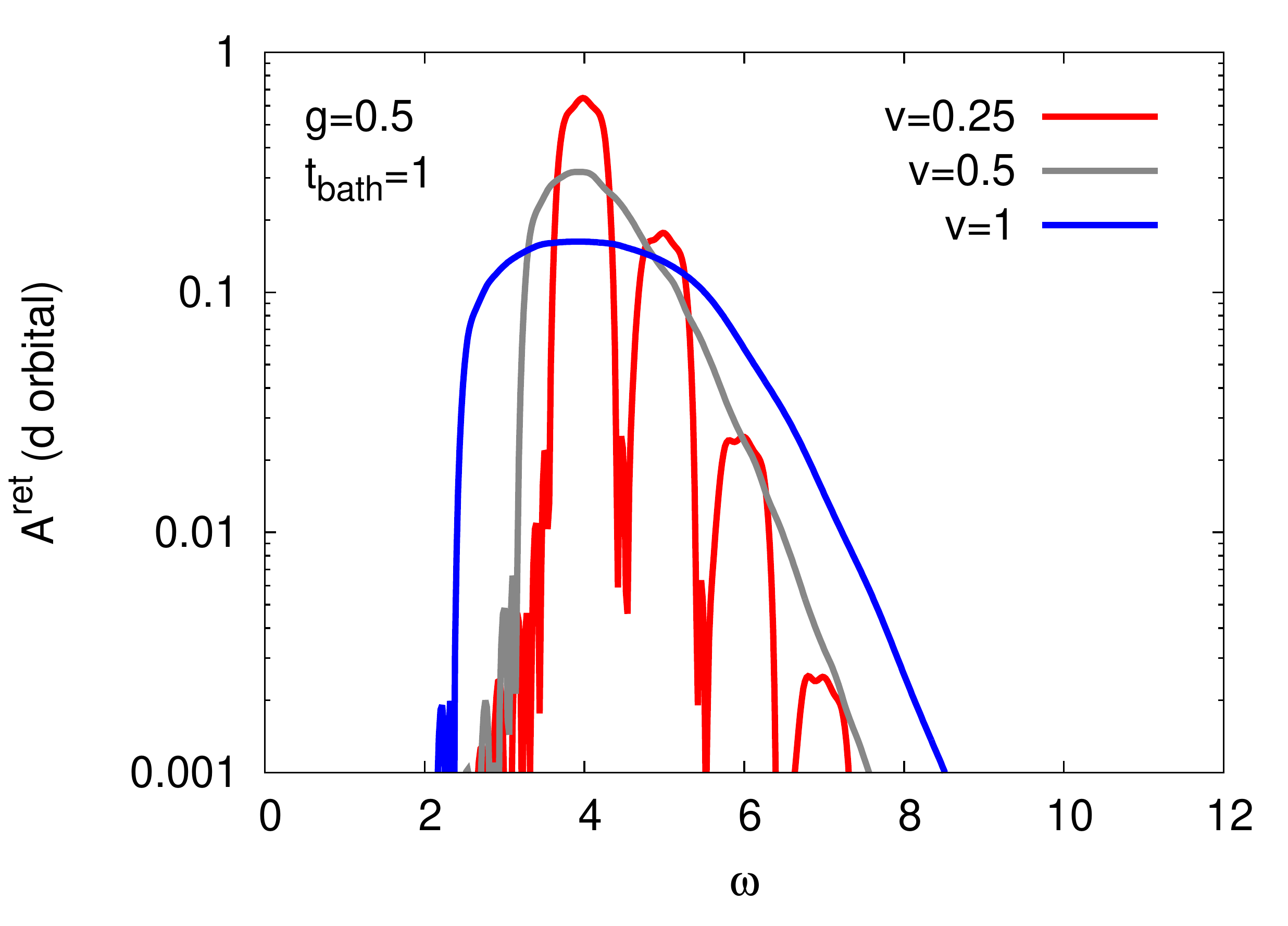} 
\caption{RIXS spectrum at $\omega_\text{in}-|E_\text{core}|=4$ in arbitrary units (a.u.) and spectral function $A^\text{ret}$ (right panels) of the equilibrium system with $g=1$ (top panels) and $g=0.5$ (bottom panels), for the indicated values of the $d$-electron hopping $v$ and core-bath coupling $t_\text{bath}$.}
\label{fig_eq}
\end{center}
\end{figure*}

The strategy behind the nonequilibrium DMFT based RIXS approach is to explicitly incorporate the core level into the impurity model, and to simulate the excitation of the core electrons by a sufficiently weak probe pulse $E_\text{probe}(t)=E_\text{probe}f_\text{probe}(t-t_\text{probe})\sin(\omega_\text{in}(t-t_\text{probe}))$, with amplitude $E_\text{probe}$, frequency $\omega_\text{in}$ and an envelope $f_\text{probe}(t-t_\text{probe})$ centered at time $t_\text{probe}$, and to measure the correlation functions $-i\langle T_\mathcal{C}P_\sigma(t)P^\dagger_{\sigma'}(t')\rangle$ (with $P_\sigma=c^\dagger_\sigma d_\sigma$), which correspond to self-energies of the outgoing photons. This allows to compute the number $N_\text{photon}(t)$ of photons emitted during and after the probe pulse, for arbitrary energy $\omega_\text{out}$.\cite{Eckstein2021} The RIXS signal is given by $\lim_{t\rightarrow\infty} N_\text{photon}(t)$. 
We use here the non-crossing approximation (NCA) as impurity solver.\cite{Keiter1971,Eckstein2010} This method treats all the hybridization-expansion diagrams without crossing hybridization lines. 

Since the phonons couple locally to the total charge $n_d+n_c$ on a given site, we can use a Lang-Firsov transformation\cite{Lang1962} to decouple the electrons and phonons, and obtain a description of $H$ in terms of polaron operators. Integrating out the phonons then yields a dressing of the hybridization-expansion diagrams\cite{Werner2007,Werner2013} in which all the creation and annihilation operators $d_\sigma^{(\dagger)}$ associated with the hybridization function (which change the occupation on the site) are connected by bosonic functions with the explicit form
\begin{align}
&K(t,t')=\exp\Bigg[
-\frac{g^2}{\omega_0^2}\frac{s^>s^<}{\sinh(\beta\omega_0/2)} \nonumber\\
&\times \big\{ \cosh((\beta/2-i(t^>-t^<))\omega_0) - \cosh(\beta\omega_0/2) \big\}
\Bigg],
\end{align}
where the greater (lesser) sign refers to the first (second) argument on the Kadanoff-Baym contour,\cite{Aoki2014} and $s$ is $+1$ ($-1$) for creation (annihilation) operators. In addition, the interactions $U$, $U_c$ and $U_{cd}$ are reduced by $\tfrac{2g^2}{\omega_0}$. 
In the NCA treatment of the electron-phonon coupling introduced in Ref.~\onlinecite{Werner2013}, each hybridization function $\Lambda_\sigma(t,t')$ is multiplied with $K(t,t')$ (where $s^>s^<=-1$), and also the physical Green's functions $G_{d,\sigma}$ and $G_{c,\sigma}$ are multiplied in a similar manner. This treatment is expected to give qualitatively correct results in the Mott insulating regime, and we will use it in the following calculations. 

The core bath is a free-electron bath with a box-shaped density of states $\rho_\text{bath}$ in the energy range $-4 < \epsilon - E_\text{core} < 4$, corresponding to a Green's function $G^0_{\text{bath}}(t,t')=-i\int d\epsilon e^{-i\epsilon(t-t')}\rho_\text{bath}(\epsilon)[\theta_\mathcal{C}(t,t')-f_\beta(\epsilon)]$ ($f_\beta(\epsilon)$ is the Fermi function for inverse temperature $\beta$ and $\theta_\mathcal{C}(t,t')$ the step function on the Kadanoff-Baym contour $\mathcal{C}$),\cite{Aoki2014} and hopping amplitude $t_\text{bath}$. The corresponding hybridization function is $\Lambda_{\text{bath}}(t,t')=t_\text{bath}^2 G^0_{\text{bath}}(t,t')$, and is also multiplied with $K(t,t')$. A sketch of the DMFT impurity problem in the action representation is shown in Fig.~\ref{fig_impurity}.  

In addition to the RIXS spectrum, we also compute the spectral function from the retarded Green's function $G^\text{ret}$ using a forward-in-time Fourier integral on a time window of length $t_\text{max}$, $A^\text{ret}(\omega,t)=-\frac{1}{\pi}\text{Im}\int_t^{t+t_\text{max}}dt' e^{i\omega(t'-t)}G^\text{ret}(t',t)$.

\section{Results}
\label{sec:results}

\subsection{Parameters}

We will consider a model with $U=8.5$, $U_c=0.5$, $U_{cd}=0.5$, $g=0.5$ and $\omega_0=1$, so that the screened interactions are $U_\text{scr}=U-\tfrac{2g^2}{\omega_0}=8$,  $U_{c,\text{scr}}=0$ and $U_{cd,\text{scr}}=0$. In realistic systems, there should be a nonzero core-valence interaction, which leads to an energy shift of the RIXS spectra, but this effect is not our main interest in this study. The level splitting is chosen as $\Delta=10$, which is much smaller than typical core-valence energy splittings, but large enough that the simulations are representative of photo-excitations from completely filled core levels. With this choice of parameters, the equilibrium system is half-filled ($n_d=1$) for $\mu=9$, the core level is approximately at energy $E_\text{core}=-14$, while the upper and lower Hubbard bands in the Mott regime are near energy $\pm 4$. If $g$ is large enough compared to the bandwidth $4v(0)$, these Hubbard bands split into phonon sidebands with an energy separation of $\omega_0$. 

For hopping $v(0)=0.25$, the chosen ratios between the interaction parameters, phonon frequency and bandwidth are approximately in line with those of the organic Mott insulator ET-F2TCNQ.\cite{Singla2015} We will however consider different bandwidths $v(0)$ in this study, since this has a significant effect on the mobility of the $d$ electrons, and hence the phonon excitations. 

In our RIXS measurement, we will consider probe pulses $f_\text{probe}=E_\text{probe}\exp(-[(t-t_\text{probe})/2]^2)$ with a Gaussian envelope and $E_\text{probe}=0.05$. This field is sufficiently small that the measured signal is quadratic in $E_\text{probe}$.  In the following discussion, we use the phonon frequency, $\omega_0=1$, as the unit of energy ($\hbar/\omega_0$ as the unit of time).\footnote{In the case of ET-F2TCNQ, $\omega_0\approx 0.1$ eV.}  The inverse temperature of the initial equilibrium state is $\beta=5$.

\subsection{Equilibrium results}
\label{sec:results:eq}

We start with a brief analysis of the $d$-electron spectral function $A^\text{ret}(\omega)$, which is shown in the right panels of Fig.~\ref{fig_eq} for hoppings $v=1$, $0.5$ and $0.25$, and for the phonon couplings $g=1$ (top panels) and $g=0.5$ (bottom panels). Despite the shallow core level, these spectra are independent of the core bath, so we only show the results for $t_\text{bath}=1$. For all considered hoppings and bath couplings, the equilibrium system is clearly Mott insulating with a gap size $>4$. Because of the particle-hole symmetric situation (in the presence of a full core level), the $d$-electron spectra are symmetric around $\omega=0$, and the figure only plots the upper Hubbard band. 

For the weaker phonon coupling $g=0.5$, the Hubbard band does not exhibit clearly resolved phonon features for hopping $v=1$ and $0.5$ (see lower-right panel in Fig.~\ref{fig_eq}), but the spectral function has a broader tail on the high-energy side compared to the pure Hubbard model. The smaller the hopping, the smaller the width of the Hubbard band, while the energy remains roughly fixed at $U_\text{scr}/2=4$, as expected. For hopping $v=0.25$ the ratio $g/v$ becomes large enough that the upper Hubbard band splits into subbands separated by the phonon energy $\omega_0=1$. The dominant sub-band at energy $\omega=4$ represents doublon creation without phonon emission, while the higher energy features correspond to doublon insertion with simultaneous emission of $n=1,2,\ldots$ phonons. In the model with stronger phonon coupling $g=1$ (upper-right panel in Fig.~\ref{fig_eq}), phonon-related structures appear in the spectral function already for $v=1$, and clearly resolved sidebands are found for $v=0.5$ and $v=0.25$ up to high orders. The width of these subbands shrinks with decreasing $v$, so that in the atomic limit, the spectral function of our model exhibits a comb of sharp phonon peaks.  
This behavior is in stark contrast to the RIXS signal discussed below, which does not exhibit any phonon features in the atomic limit (and without core bath), since the excitation and filling of the core hole does not change the occupation on the given site and hence does not couple to the phonons.  

\begin{figure}[t]
\begin{center}
\includegraphics[angle=0, width=0.47\columnwidth]{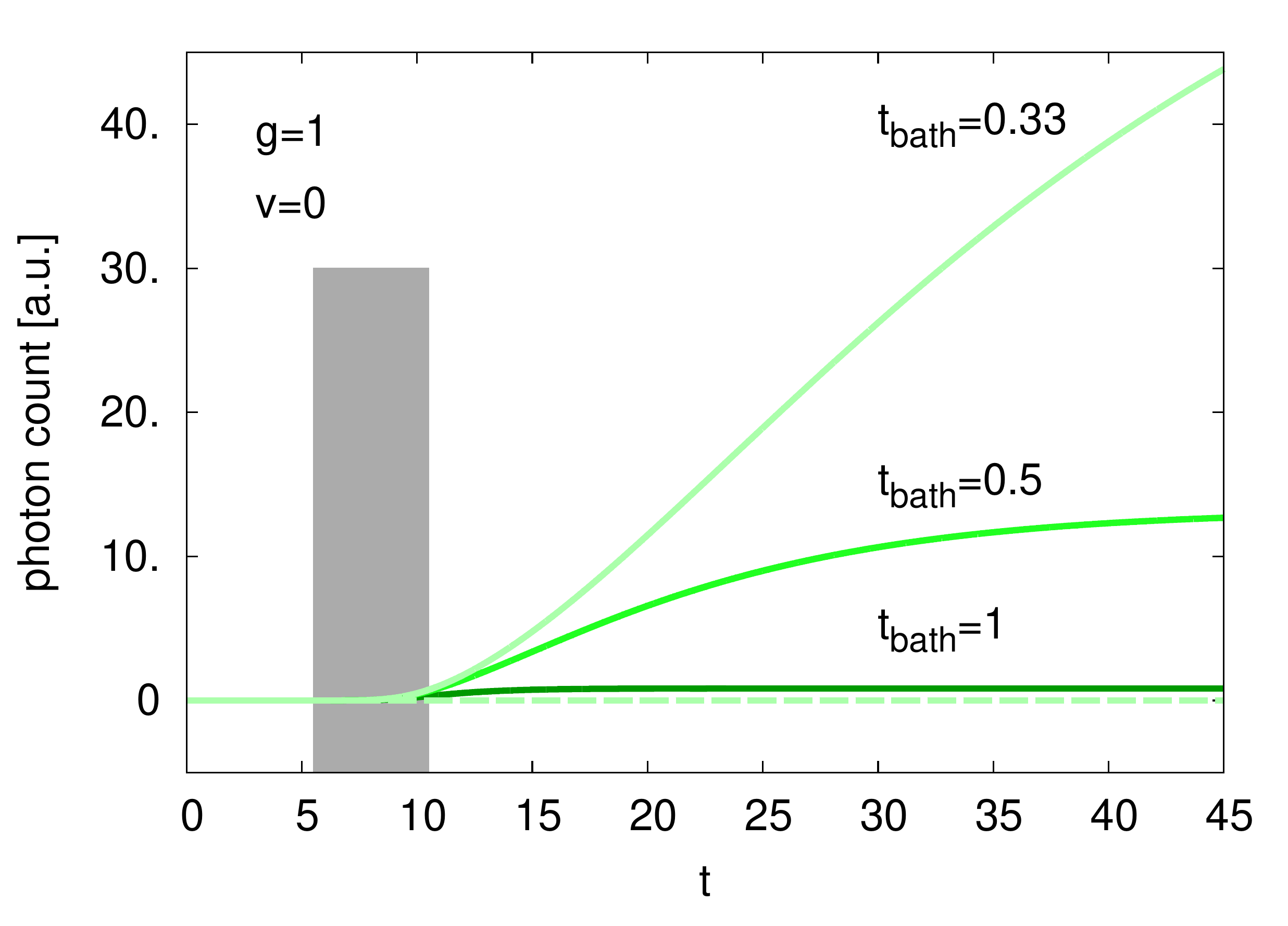} 
\hfill
\includegraphics[angle=0, width=0.47\columnwidth]{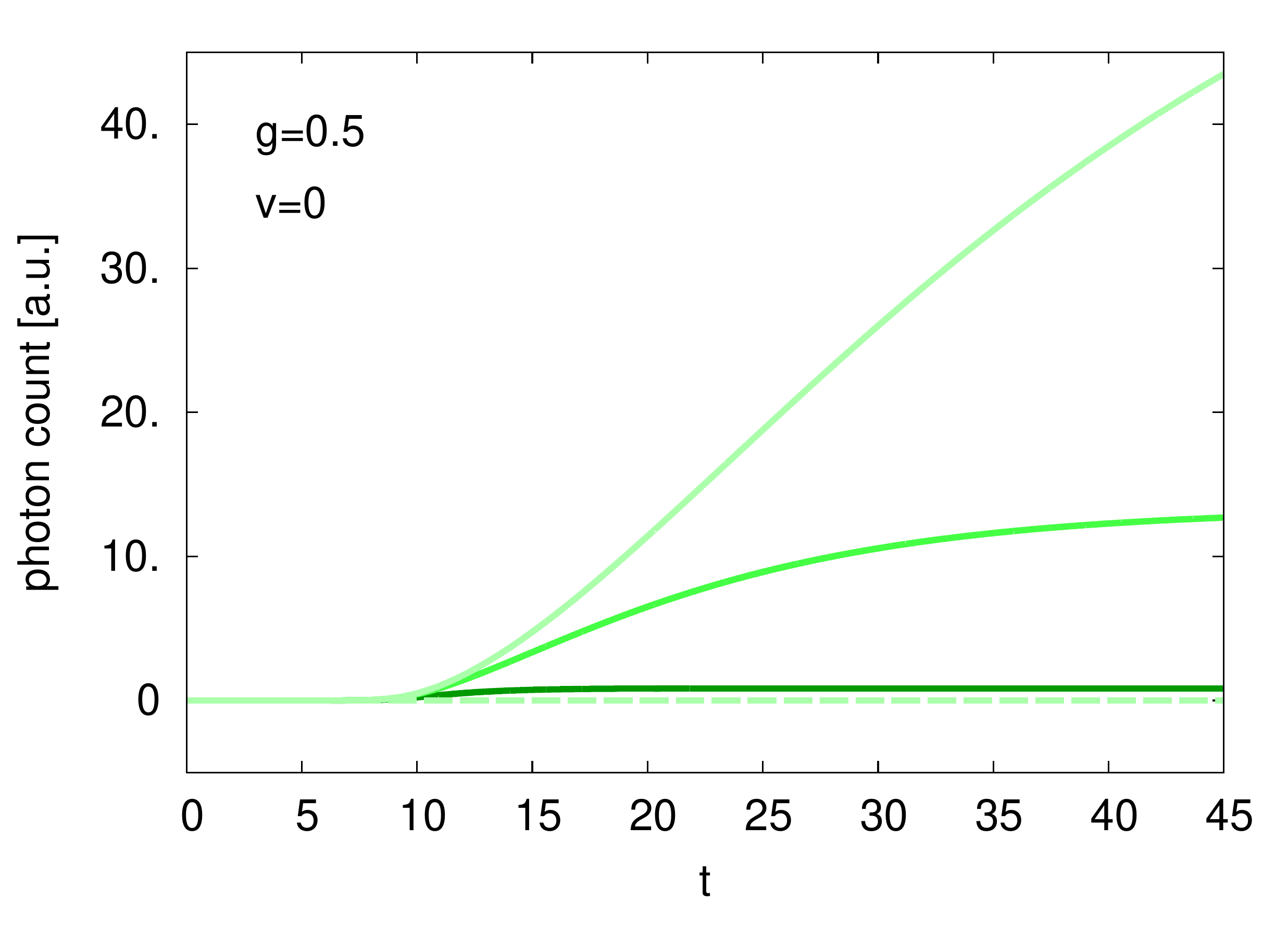} 
\hfill
\includegraphics[angle=0, width=0.47\columnwidth]{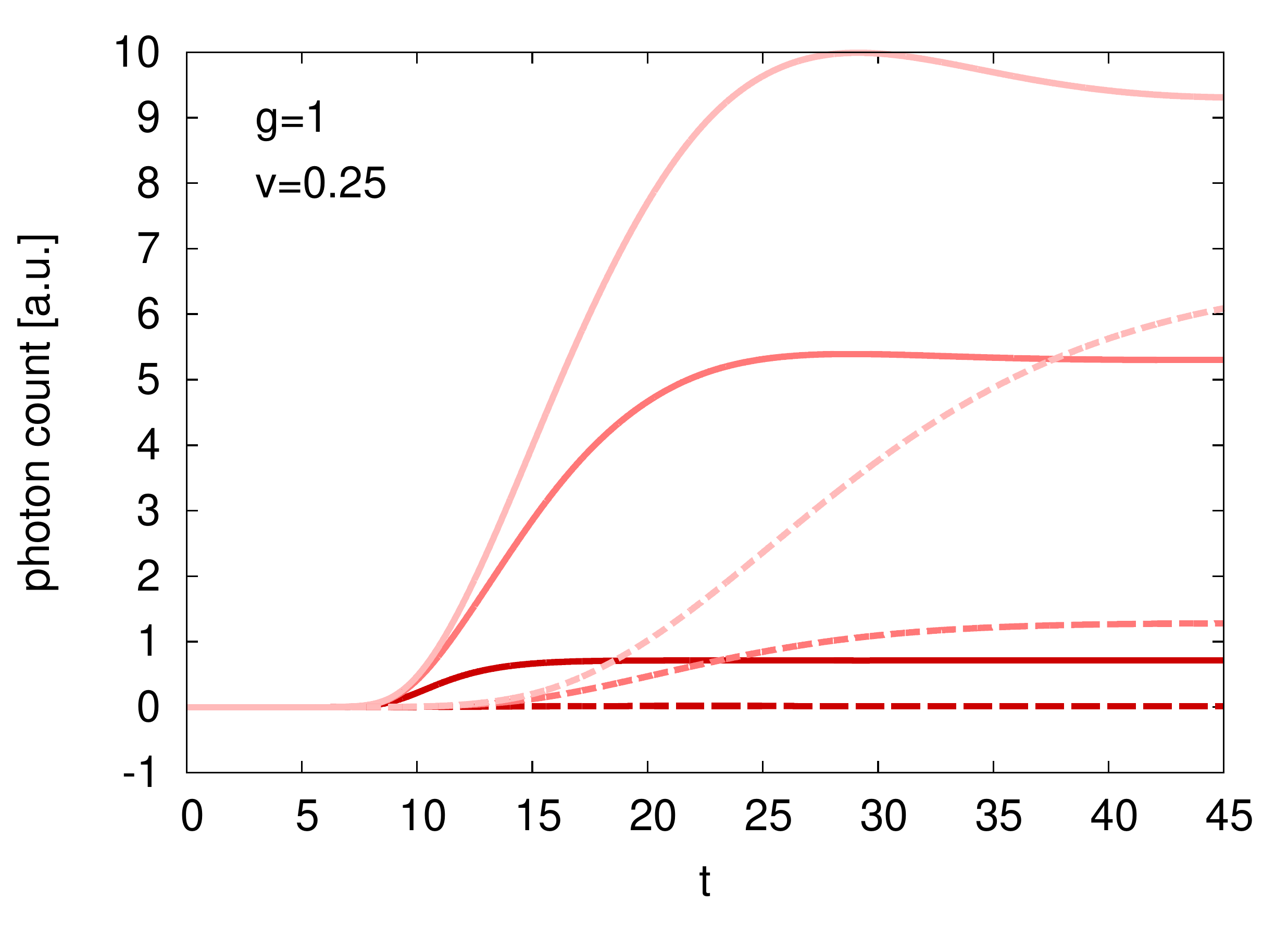} 
\hfill
\includegraphics[angle=0, width=0.47\columnwidth]{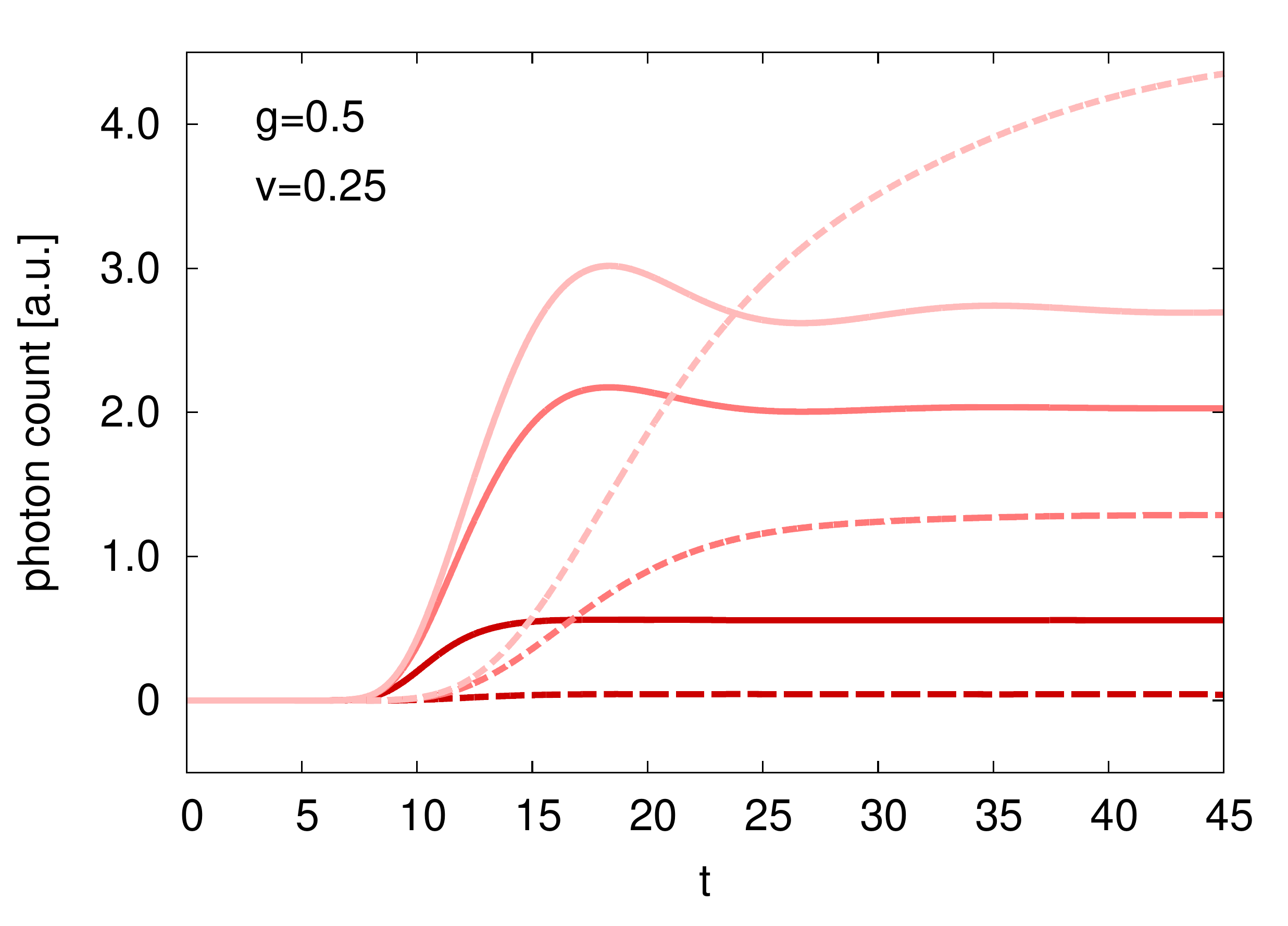} 
\hfill
\includegraphics[angle=0, width=0.47\columnwidth]{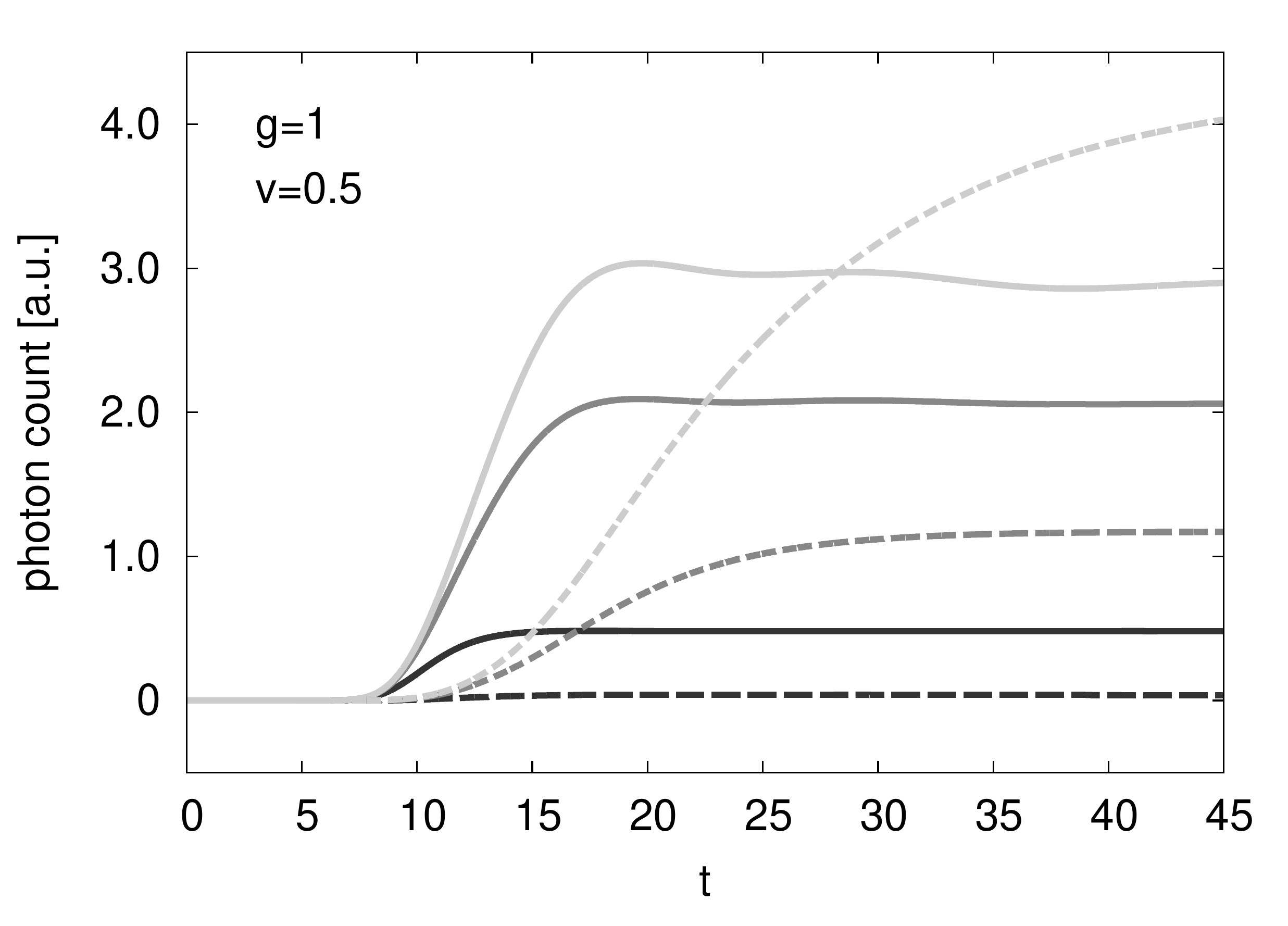} 
\hfill
\includegraphics[angle=0, width=0.47\columnwidth]{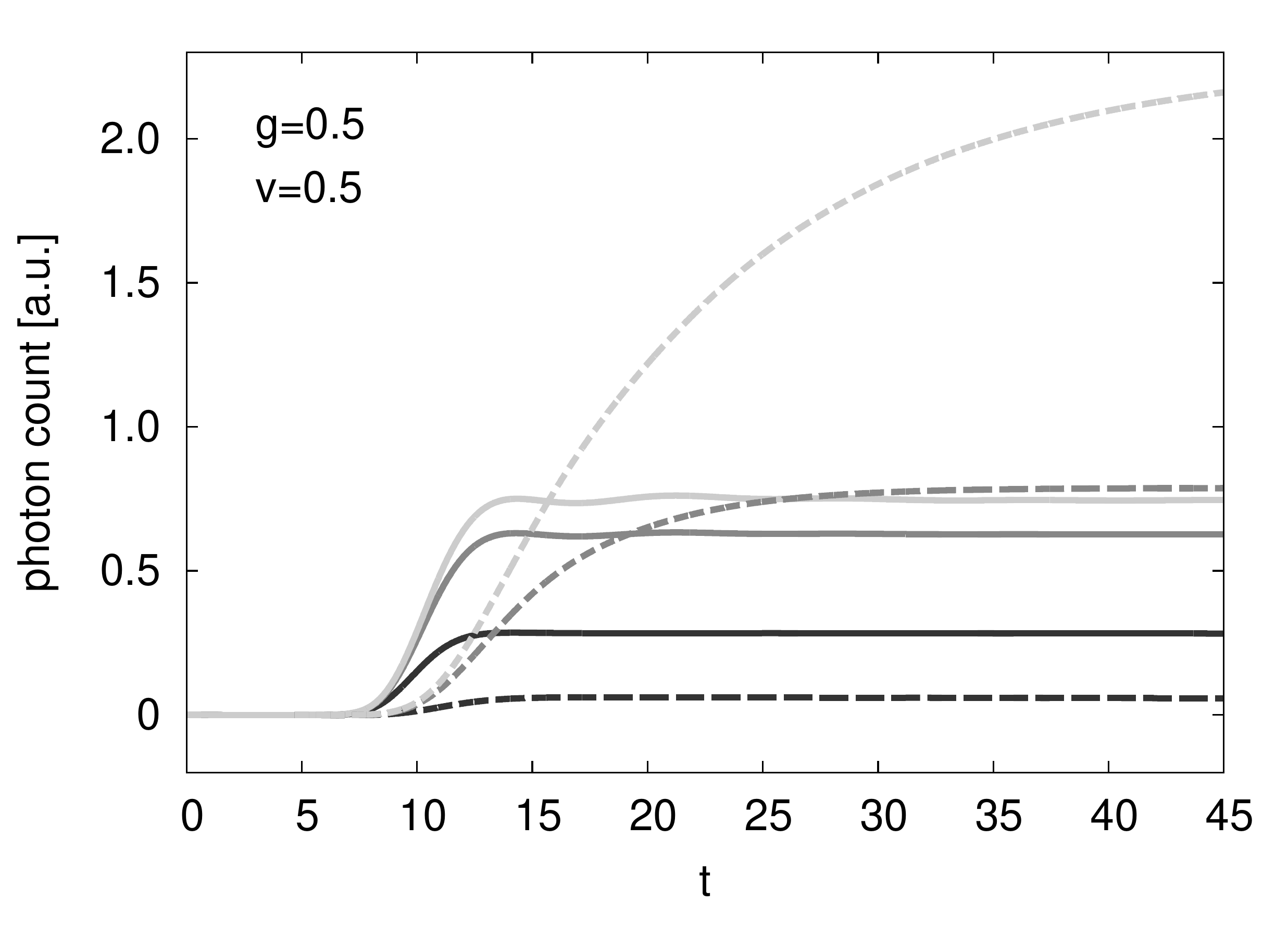} 
\hfill
\includegraphics[angle=0, width=0.47\columnwidth]{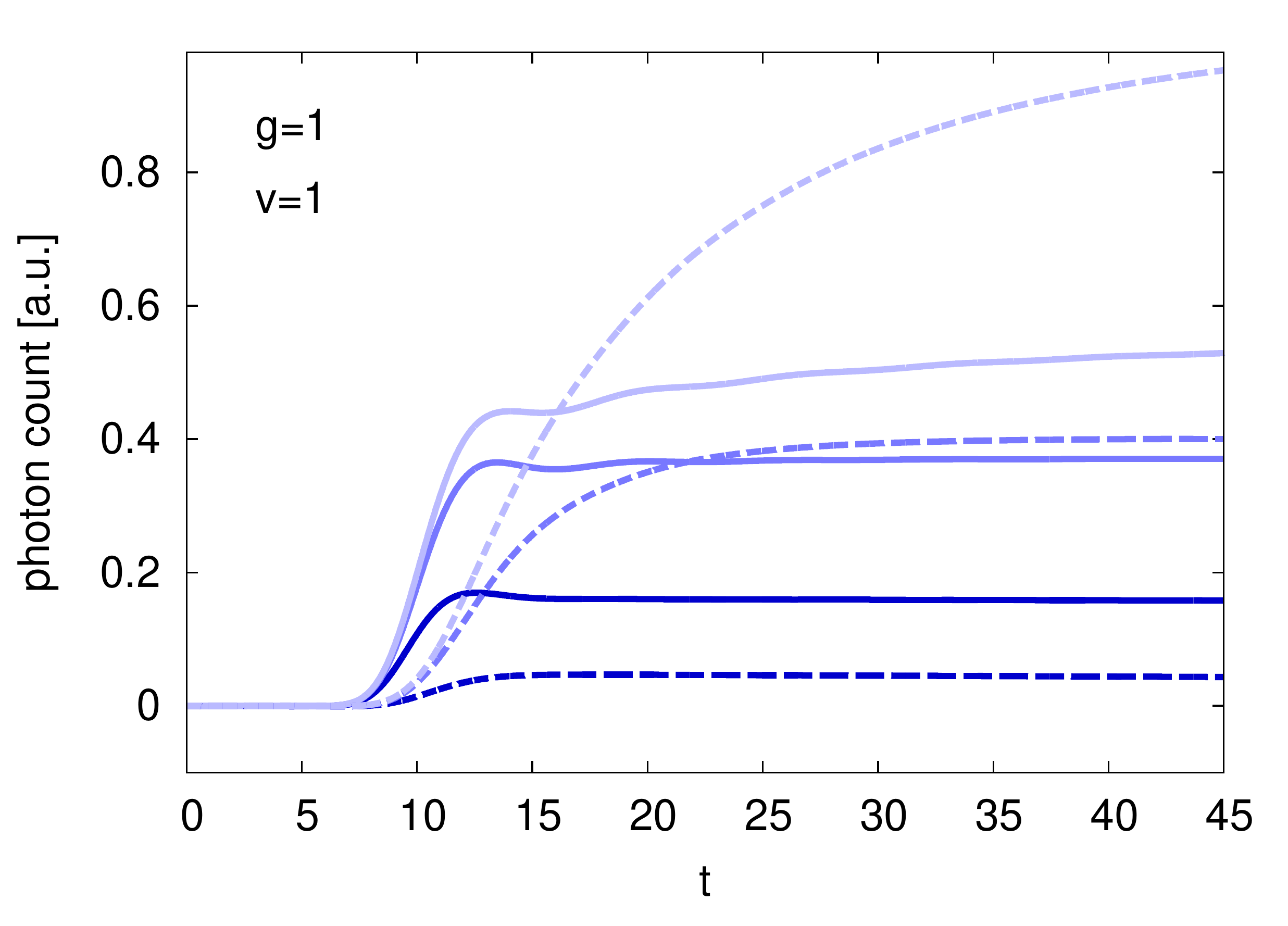} 
\hfill
\includegraphics[angle=0, width=0.47\columnwidth]{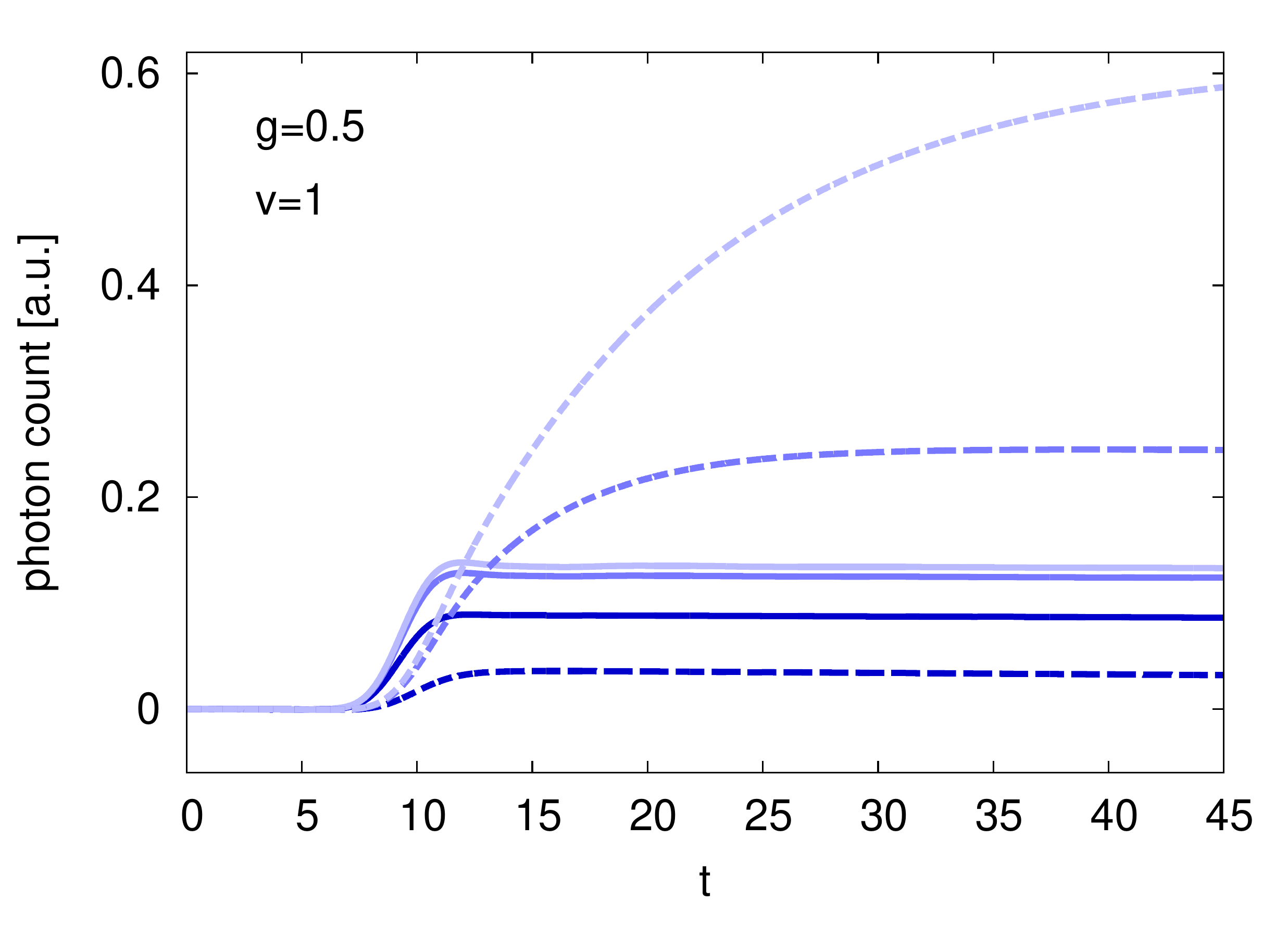} 
\caption{Total photon count $N_\text{photon}(t)$ for $\omega_\text{in}-|E_\text{core}|=4$ and indicated values of the hopping $v$. Solid (dashed) lines show the results for $\omega_\text{out}-|E_\text{core}|=4$ ($-4$), and the dark, intermediate and light colors correspond to $t_\text{bath}=1$, $0.5$ and $0.25$. The left panels are for $g=1$ and the right panels for $g=0.5$. The RIXS spectrum corresponds to the total photon count produced by the probe pulse, which acts in the time interval indicated by the gray shaded region in the top left panel.}
\label{fig_photoncount}
\end{center}
\end{figure}

As briefly explained in Sec.~\ref{sec:method}, and in more detail in Ref.~\onlinecite{Eckstein2021}, the RIXS signal is measured as the total number of photons emitted by the system in response to the RIXS pulse. As an illustration, the panels in Fig.~\ref{fig_photoncount} plot, for a probe pulse centered near $t_\text{probe}=8$, the time evolution of the photon count $N_\text{photon}$ at the outgoing frequency $\omega_\text{out}-|E_\text{core}|=4$, corresponding to the elastic peak (solid lines), and $\omega_\text{out}-|E_\text{core}|=-4$, corresponding to the $d$-$d$ excitation feature (dashed lines). The top panels show the results in the atomic limit $v=0$, and the other panels for $v=0.25$, $0.5$ and $1$, while the different line colors correspond to different $t_\text{bath}$. The grey shaded time interval in the top left panel indicates the approximate duration of the RIXS excitation pulse. 
A stronger core-bath coupling leads to a shorter core-hole lifetime and suppresses the probe-pulse induced coherences between the core and valence orbitals. Similarly, a larger $d$-hopping amplitude $v$ destroys the coherence and leads to a faster saturation of the photon count. A rough estimate of the core hole lifetime from the simulations for $v=0$ (atomic limit) is $\tau_\text{core}=25,14,5$ for the different core-bath coupling strengths considered. In practice, we estimate the RIXS signals from the photon count at the longest simulation time $t=45$. 

The equilibrium RIXS spectra for probe frequency $\omega_\text{in}-|E_\text{core}|=4$ are shown in the left three panels of Fig.~\ref{fig_eq} and exhibit two main features: (i) a relatively sharp peak at $\omega_\text{out}\approx\omega_\text{in}$ and (ii) a broader feature below $\omega_\text{out}-|E_\text{core}| \lesssim -4$  with a width that approximately matches the width of the Hubbard band in $A^\text{ret}(\omega)$. Feature (i) is the elastic peak, which is associated with the excitation of a core electron to an initially half-filled $d$ orbital (creation of a doublon), and the de-excitation of this electron back to the core. Feature (ii) is the $d$-$d$ charge excitation peak, which corresponds to the creation of a doublon by the core-hole excitation, the hopping of the doublon to a neighboring site, and the de-excitation from a singly occupied $d$ state. This process leaves behind a doublon-holon pair in the system, which costs an energy of $U_\text{scr}=8$. Since the elastic process does generically not involve hoppings to other sites, it essentially does not couple to phonons (within our model), which explains the absence of clear phonon features in the elastic peak. In contrast, the $d$-$d$ feature necessarily involves the hopping of a $d$ electron and a change in the total occupation of the site. This process couples to phonons, and in the equilibrium state, where the initial phonon occupation is low, it primarily results in the emission of phonons. For $g=1$, the $d$-$d$ excitation peak therefore features sidebands below the main peak near $\omega_\text{out}-|E_\text{core}| = -4$, which correspond to RIXS processes that leave behind a doublon-holon pair in the system, and additionally emit $n=1,2,\ldots$ phonons. If the ratio $g/v$ is too small for the appearance of phonon sidebands in the spectral function, we also cannot clearly resolve such phonon features in the $d$-$d$ excitation signal in RIXS. 

The effect of a shorter core-hole lifetime is not only a weaker RIXS signal, but we also notice a stronger suppression of the loss features relative to the elastic line with increasing $t_\text{core}$. This is because the doublons have less time to make excursions to the neighboring sites.  The smearing out of the phonon features is likely an artefact of the core bath coupling, and should be suppressed in the realistic case of a very deep core level. 

\begin{figure}[t]
\begin{center}
\includegraphics[angle=0, width=\columnwidth]{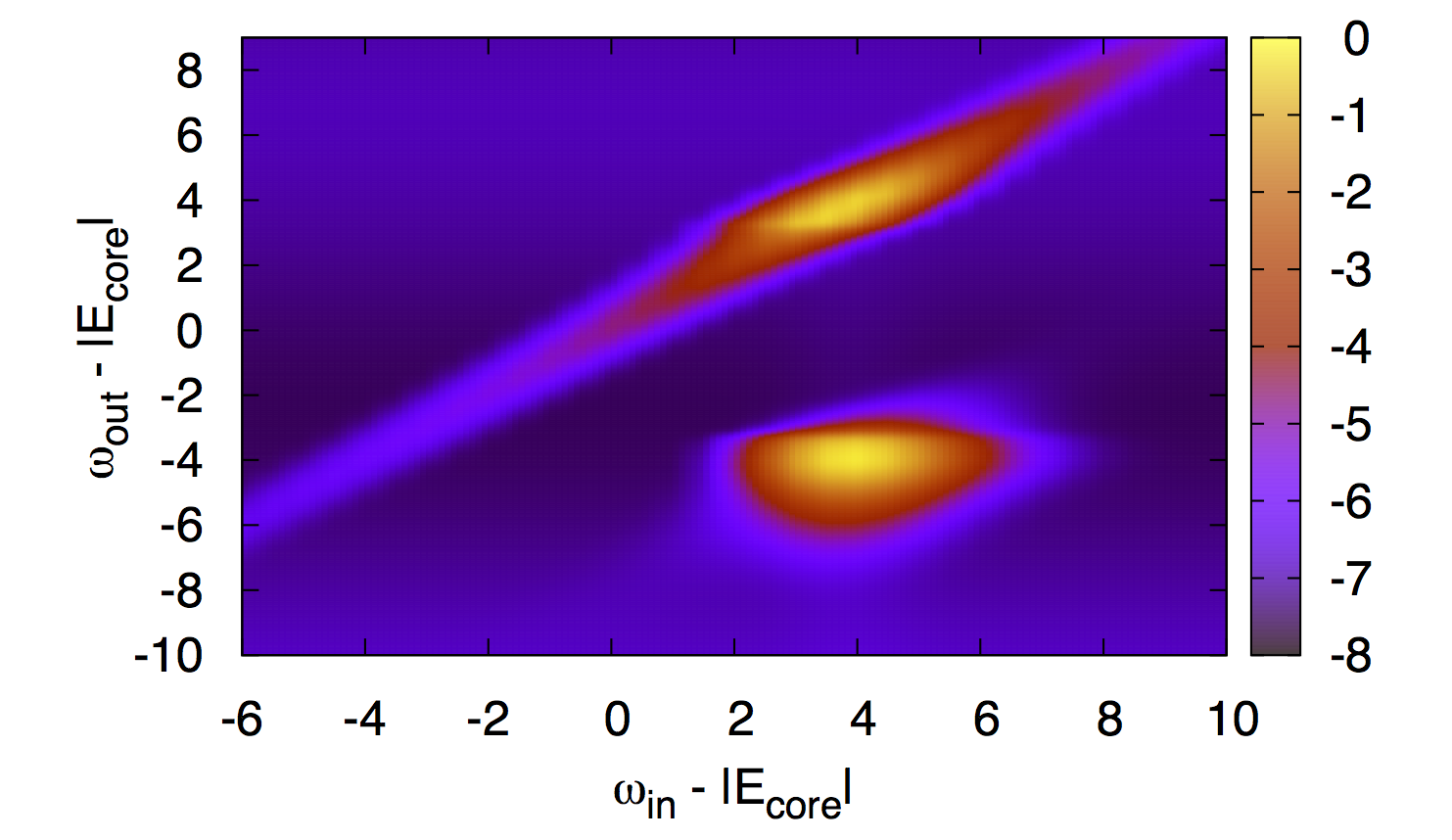} 
\includegraphics[angle=0, width=\columnwidth]{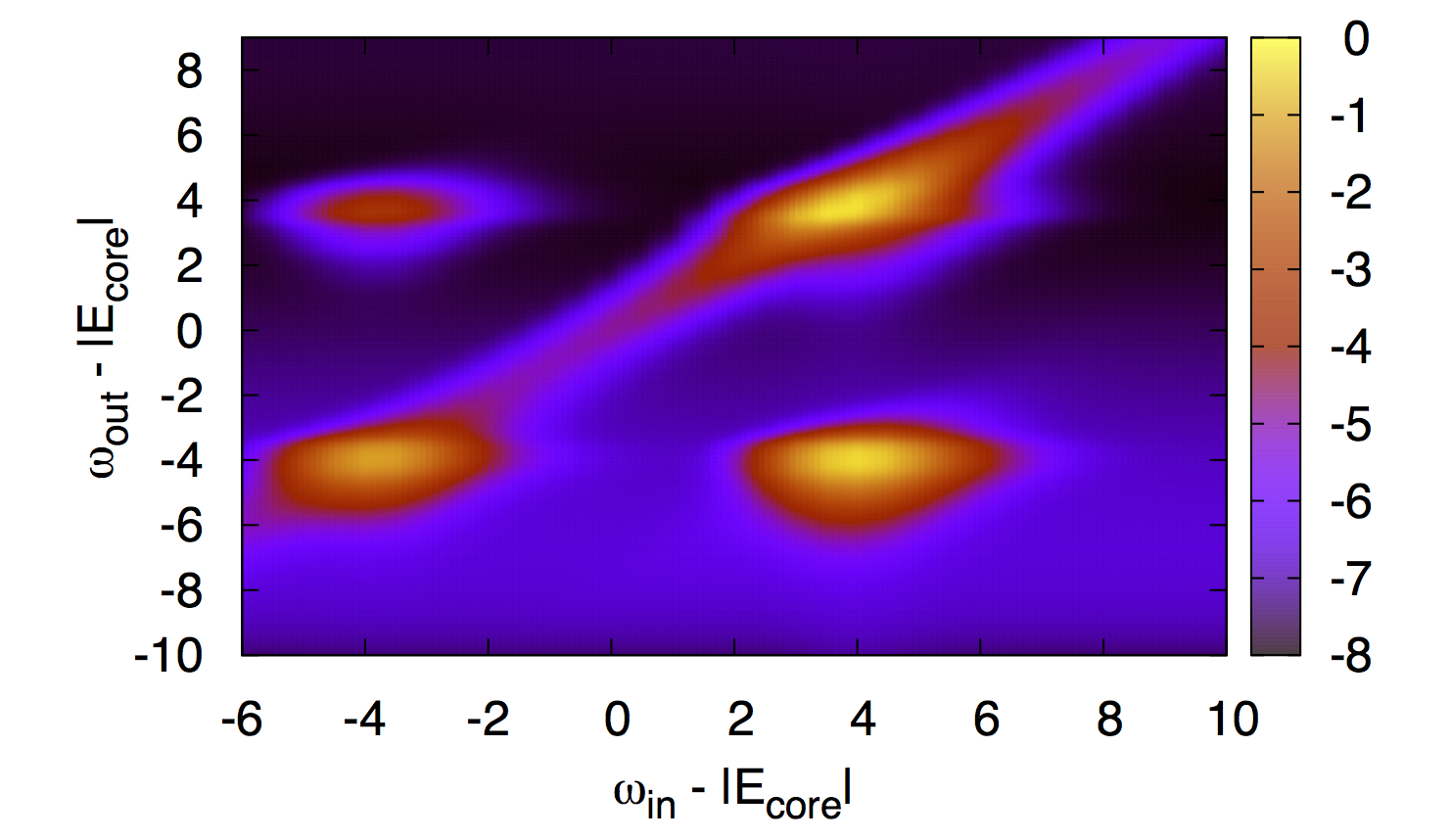} 
\includegraphics[angle=0, width=\columnwidth]{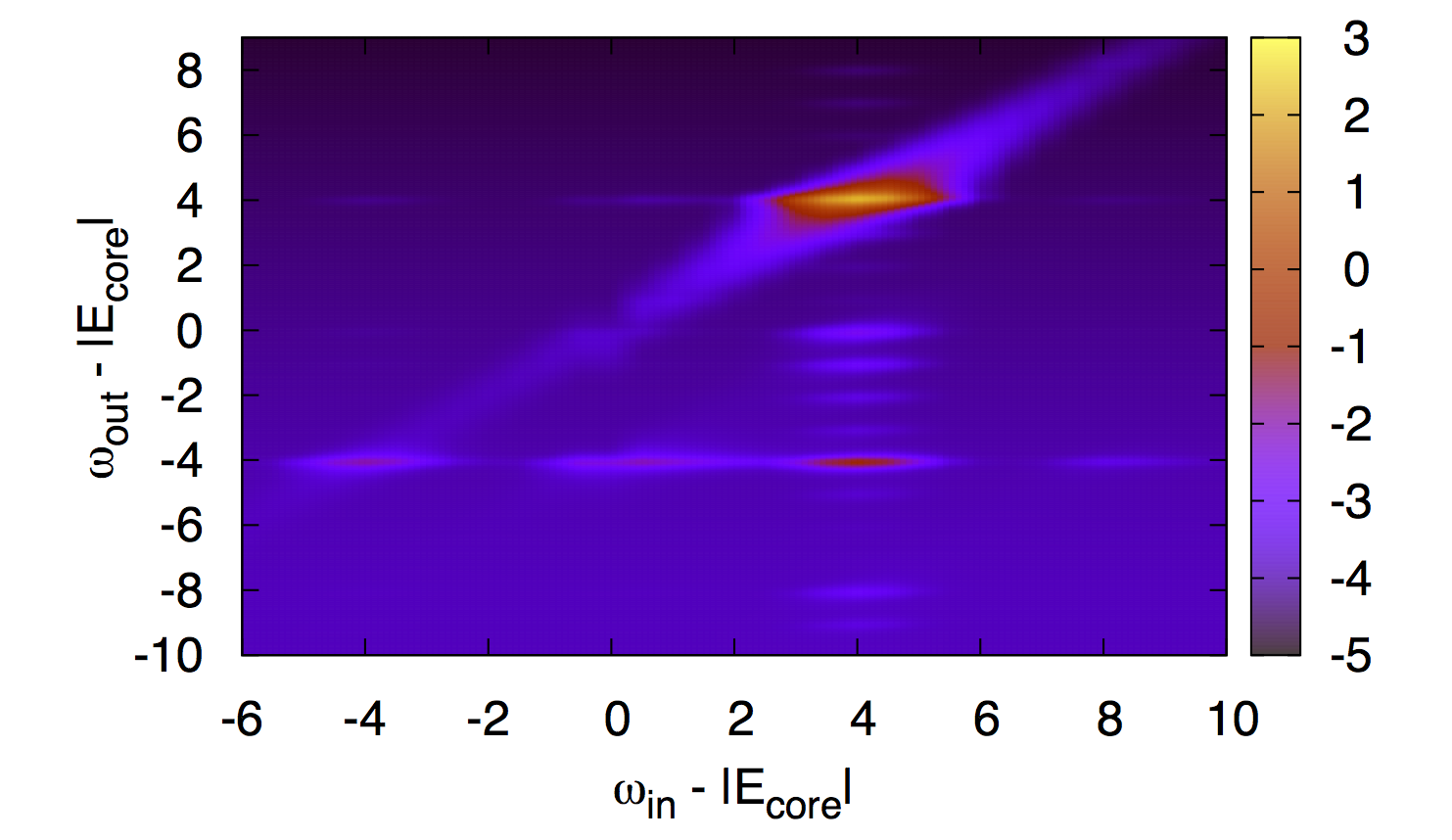} 
\caption{RIXS spectrum (log-scale plot in arbitrary units) as a function of $\omega_\text{in}$ and $\omega_\text{out}$ for the equilibrium system (top panel), photo-doped system ($\Omega_\text{pump}=8$, middle panel), and system with static electric field ($E=4$, bottom panel). The model parameters are $g=0.5$, $v=0.5$ and $t_\text{bath}=0.5$.}
\label{fig_rixs_3d}
\end{center}
\end{figure}

In the top panel of Fig.~\ref{fig_rixs_3d}, we plot the equilibrium RIXS spectrum as a function of $\omega_\text{in}$ and $\omega_\text{out}$. For later convenience, we choose here a parameter set ($g=0.5$, $v=0.5$, $t_\text{bath}=0.5$) with a small ratio $g/v$, so that phonon-related features cannot be resolved. The higher energy signal with $\omega_\text{out}\approx \omega_\text{in}$ is the elastic line, corresponding to processes such as $(\uparrow\downarrow , \uparrow) \xrightarrow{\omega_\text{in}} (\uparrow , \uparrow \downarrow) \xrightarrow{\omega_\text{out}} (\uparrow\downarrow , \uparrow)$, where the brackets represent the ($\text{core} , \text{valence}$) configuration. The signal near $\omega_\text{in}-|E_\text{core}|\approx 4$ and $\omega_\text{out}-|E_\text{core}|\approx -4$ is the $d$-$d$ excitation feature associated with doublon hopping processes, e.~g. $(\uparrow\downarrow , \uparrow) (\uparrow\downarrow , \downarrow)\xrightarrow{\omega_\text{in}} (\uparrow , \uparrow \downarrow) (\uparrow\downarrow , \downarrow)$ $ \xrightarrow{\text{hop}}(\uparrow , \downarrow) (\uparrow\downarrow , \uparrow\downarrow)   \xrightarrow{\omega_\text{out}} (\uparrow\downarrow , 0) (\uparrow\downarrow , \uparrow\downarrow)$. Here, the two brackets correspond to neighboring sites, and the core hole creation and annihilation process takes place on the first site.

Both the elastic feature and the loss feature ($d$-$d$ excitation peak) have a maximum intensity around $\omega_\text{in}-|E_\text{core}|$, and within an energy range $\Delta\omega_\text{in}\approx 4$ corresponding to the width of the upper Hubbard band. In this energy range, doublons can be efficiently created by the RIXS pulse. The horizontal elongation of the loss peak reveals its fluorescent character: kinetic energy of the photo-excited core electron can be dissipated to electronic and bosonic degrees of freedom, which reduces the energy of the emitted photons. 

\begin{figure*}[ht]
\begin{center}
\includegraphics[angle=0, width=0.49\columnwidth]{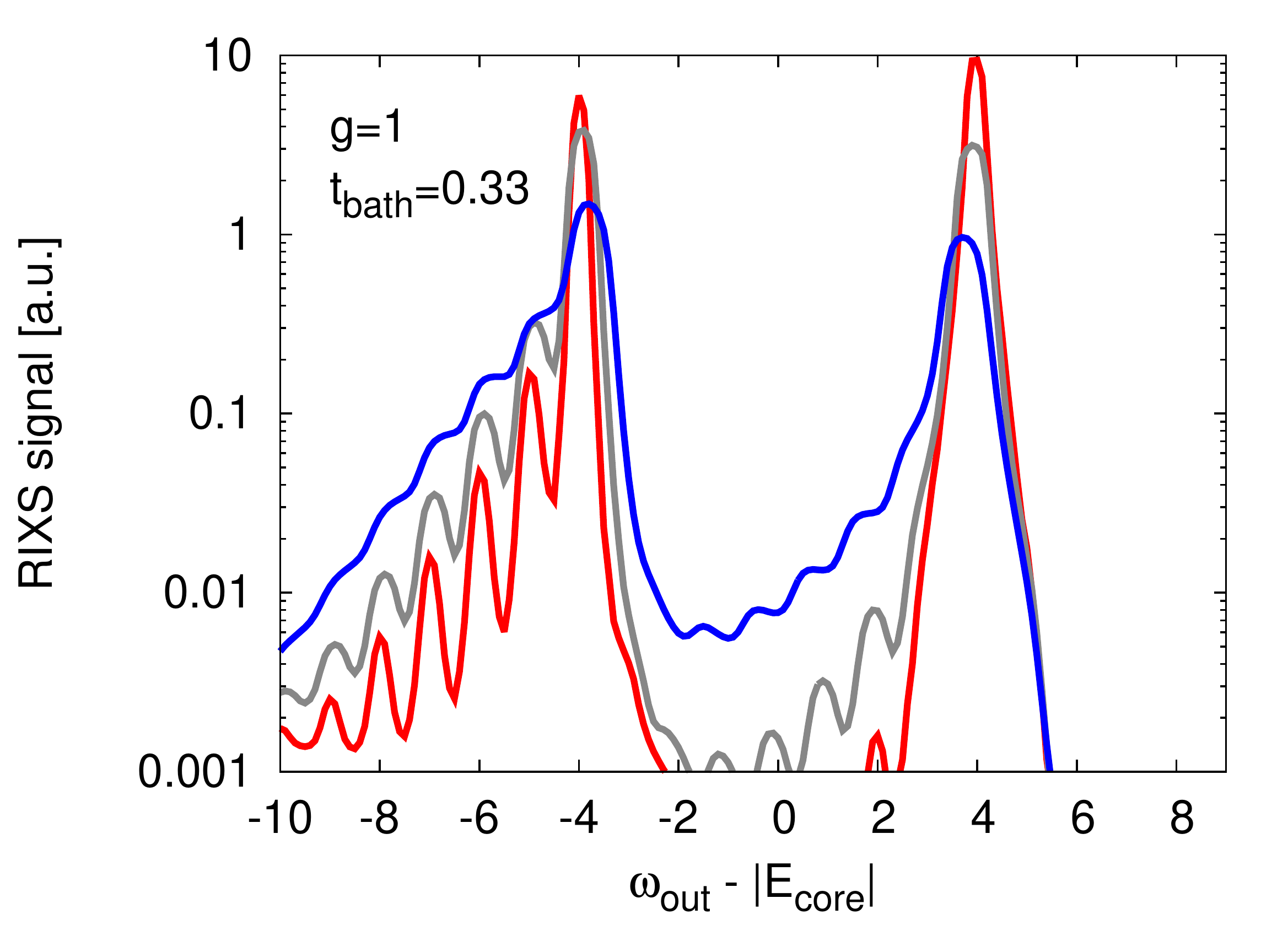}\hfill 
\includegraphics[angle=0, width=0.49\columnwidth]{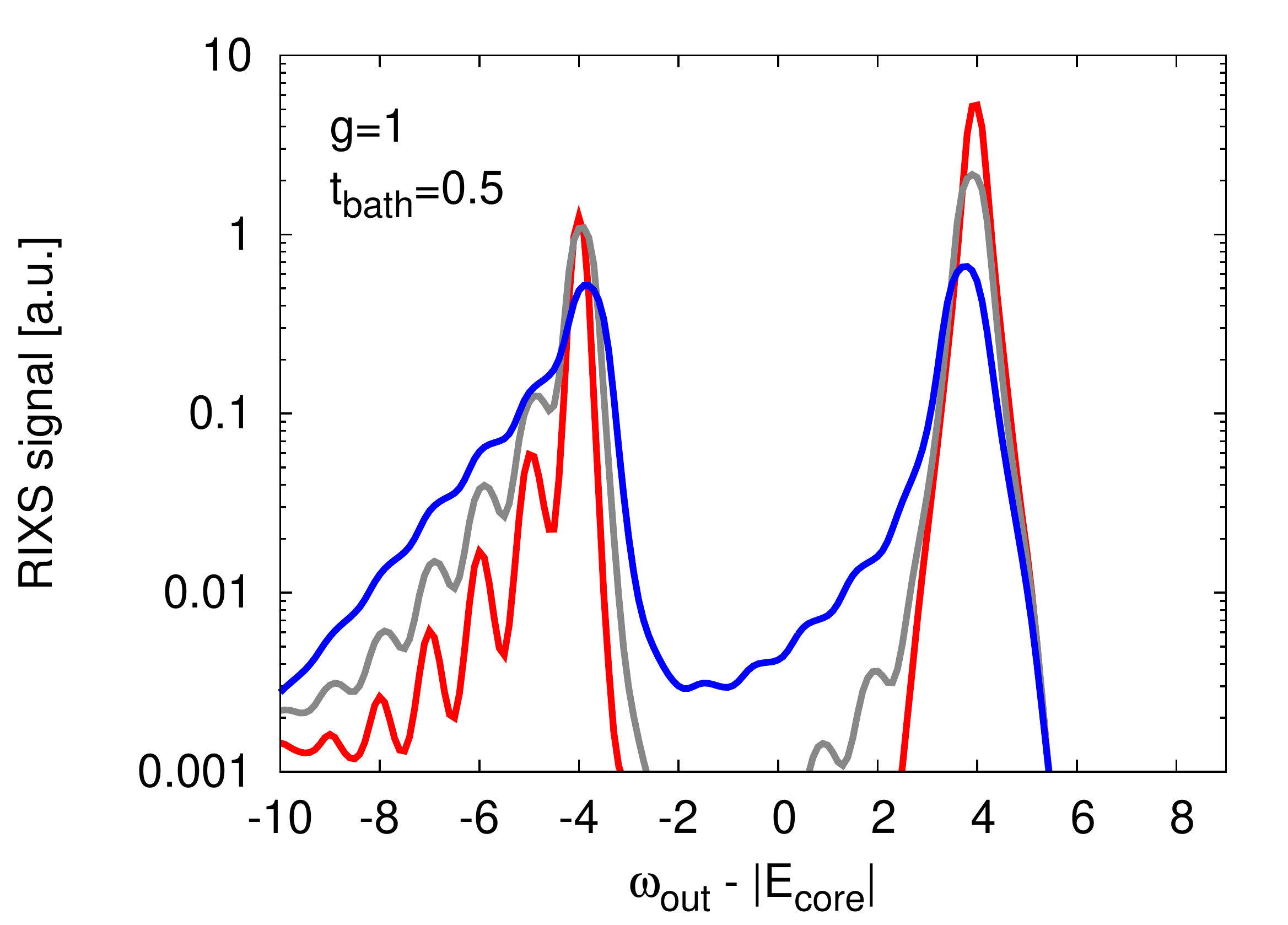}\hfill 
\includegraphics[angle=0, width=0.49\columnwidth]{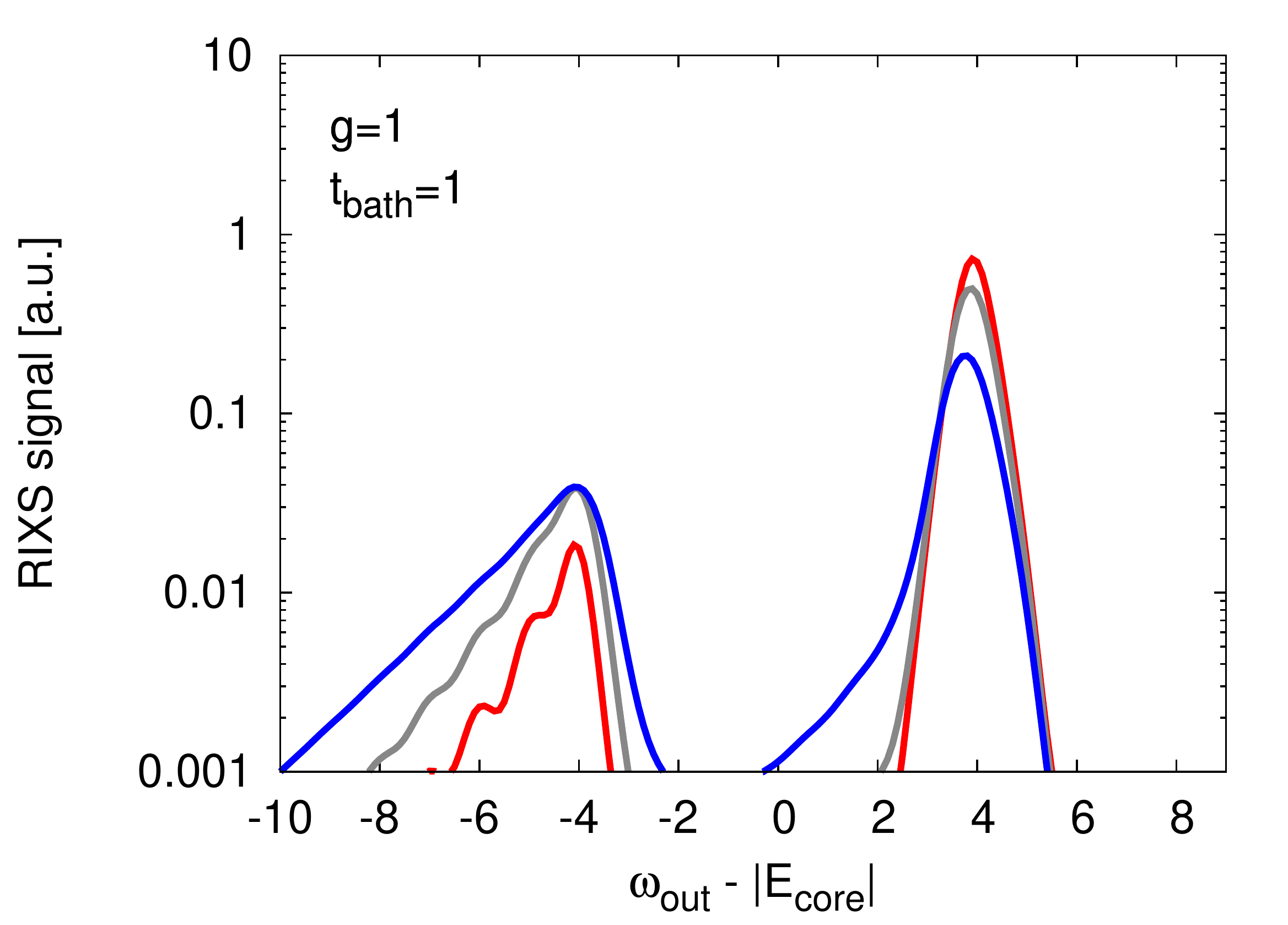}\hfill 
\includegraphics[angle=0, width=0.49\columnwidth]{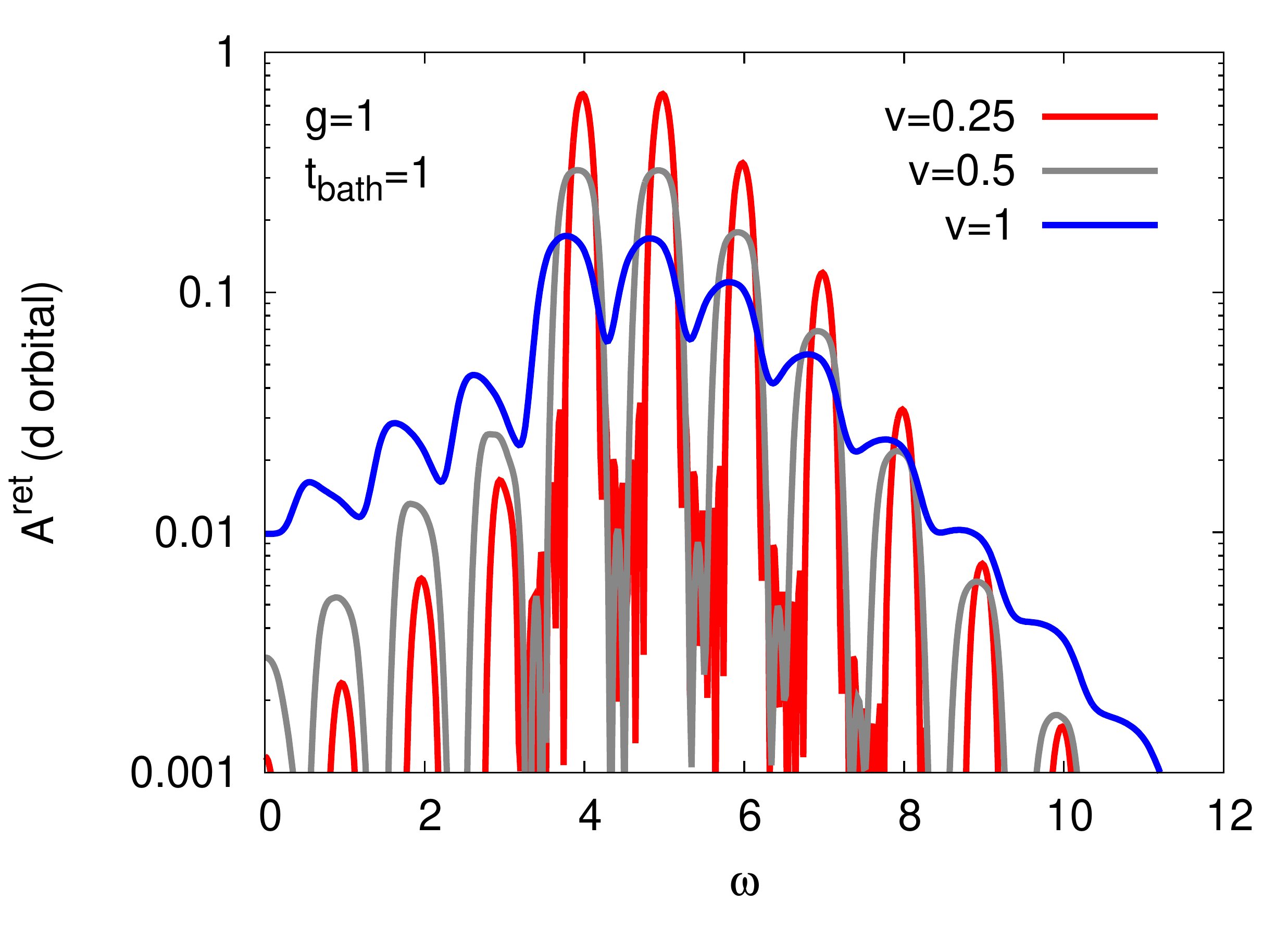}\\ 
\includegraphics[angle=0, width=0.49\columnwidth]{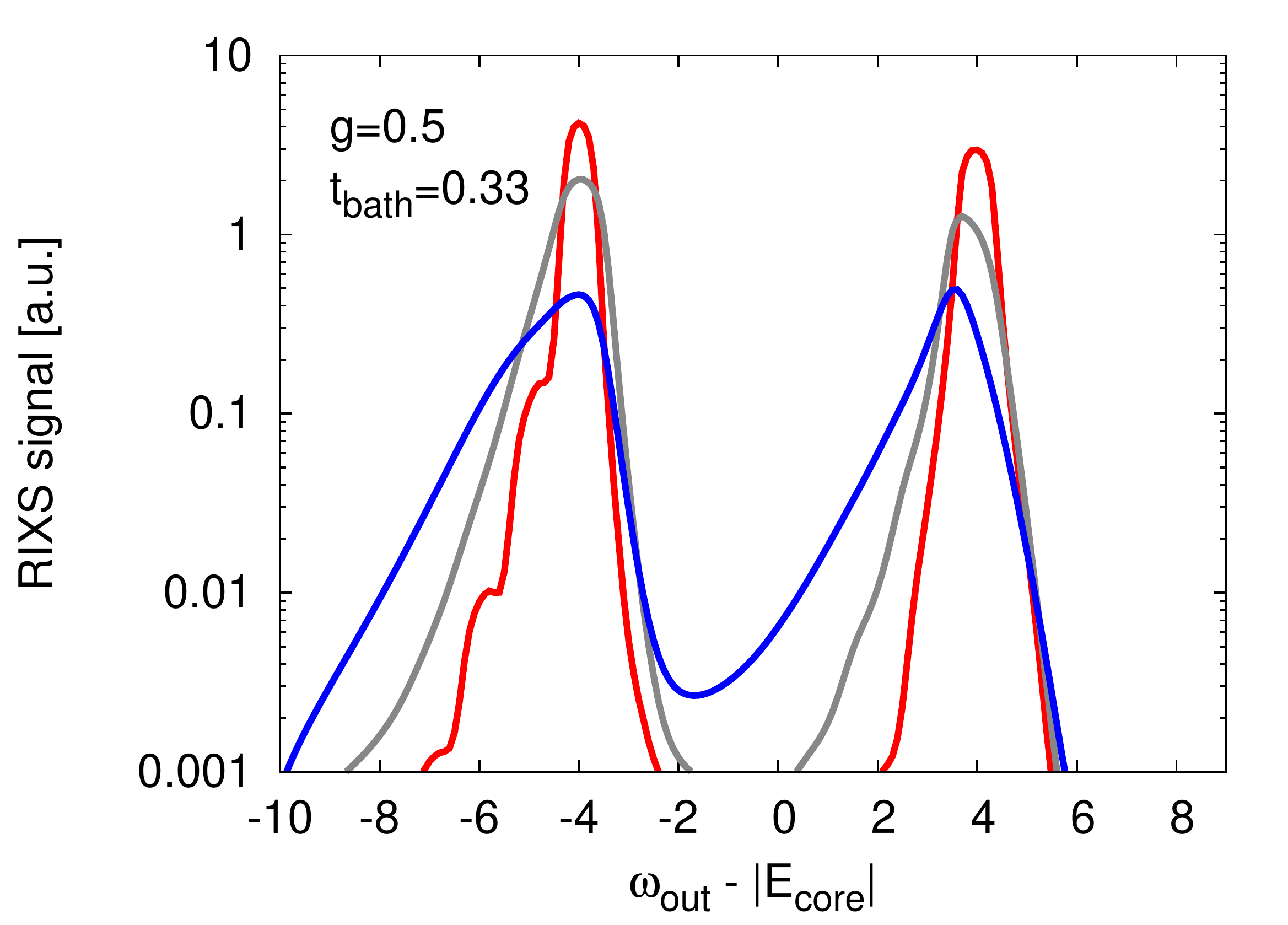}\hfill 
\includegraphics[angle=0, width=0.49\columnwidth]{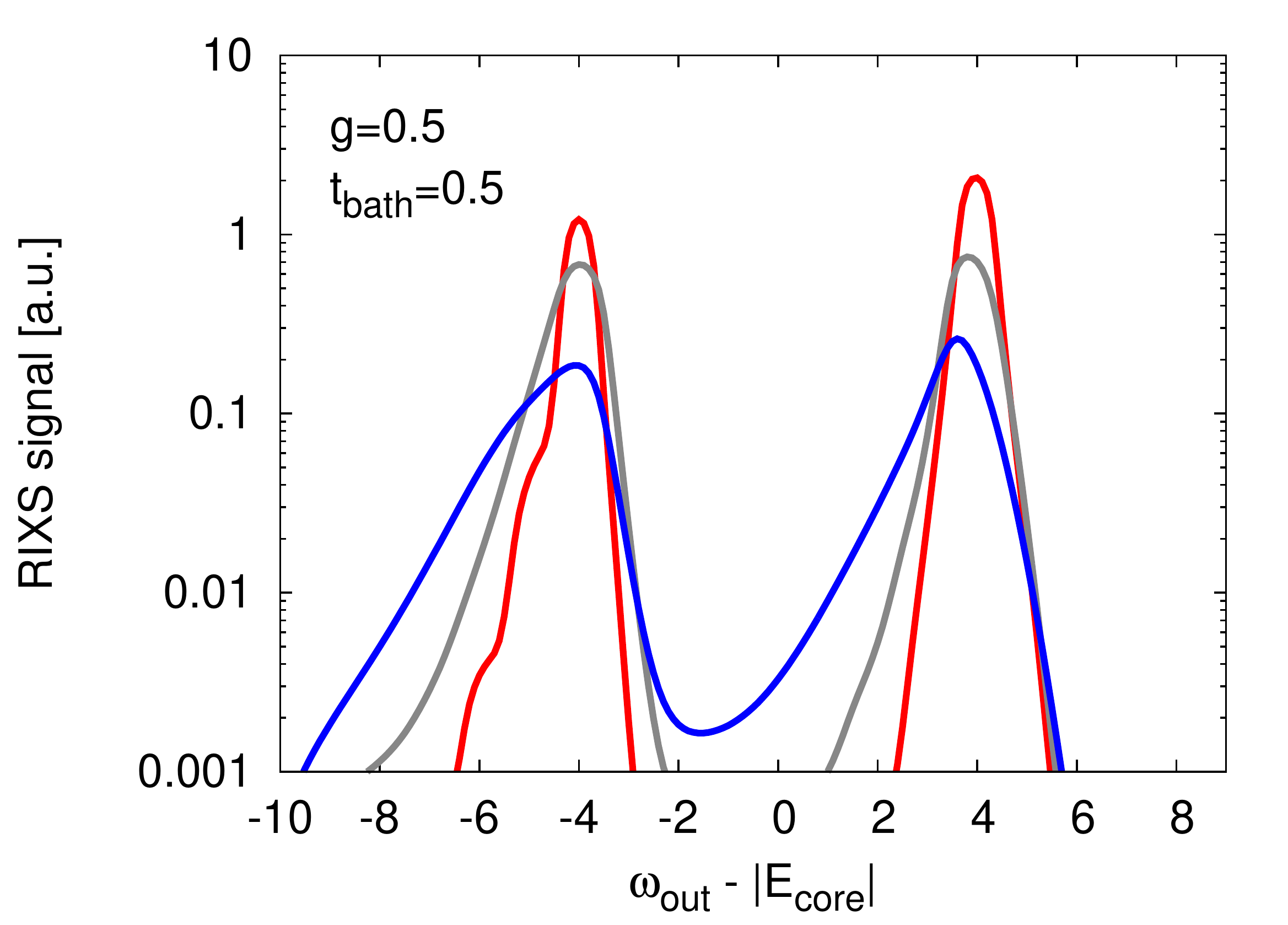}\hfill 
\includegraphics[angle=0, width=0.49\columnwidth]{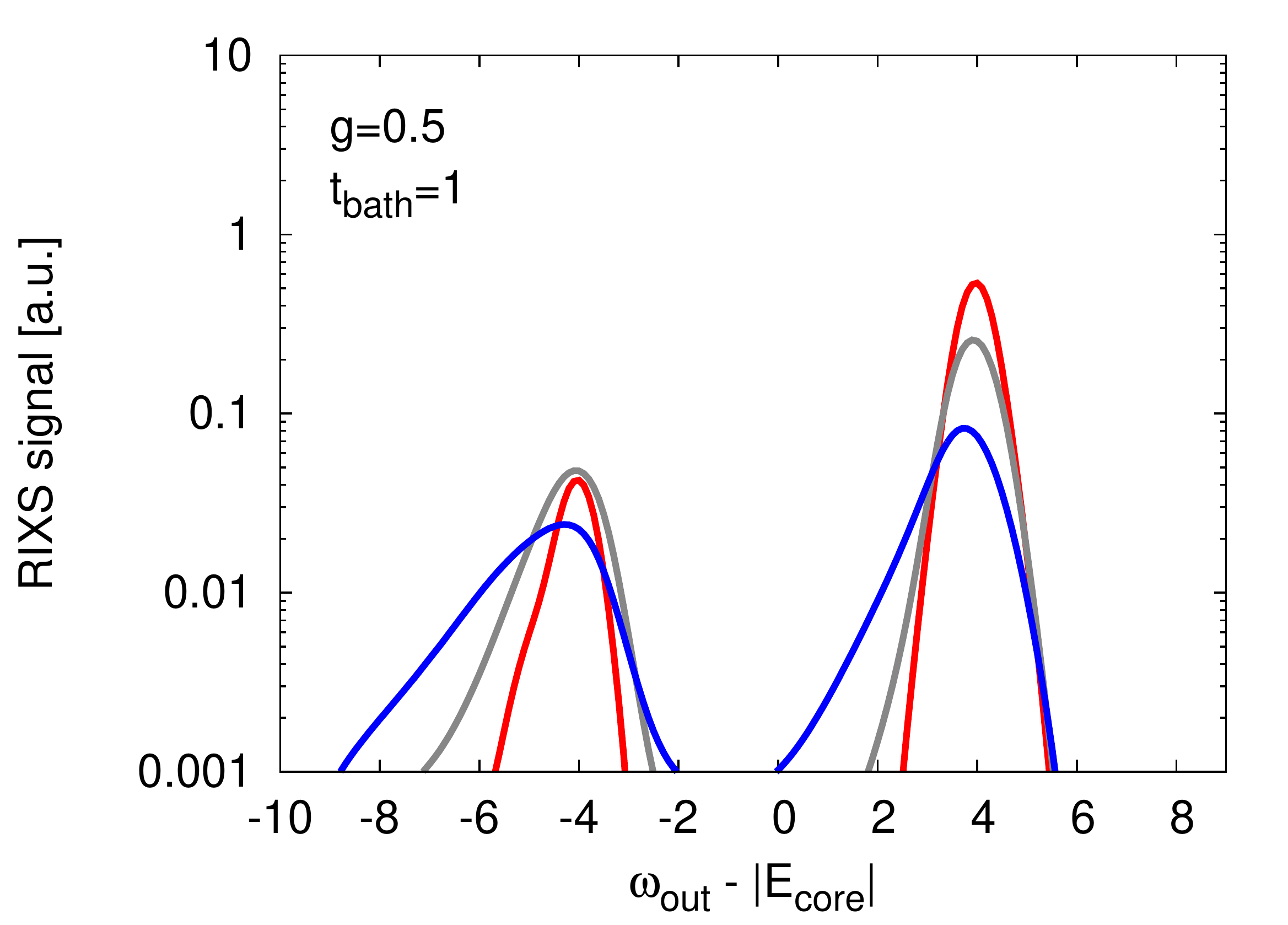}\hfill 
\includegraphics[angle=0, width=0.49\columnwidth]{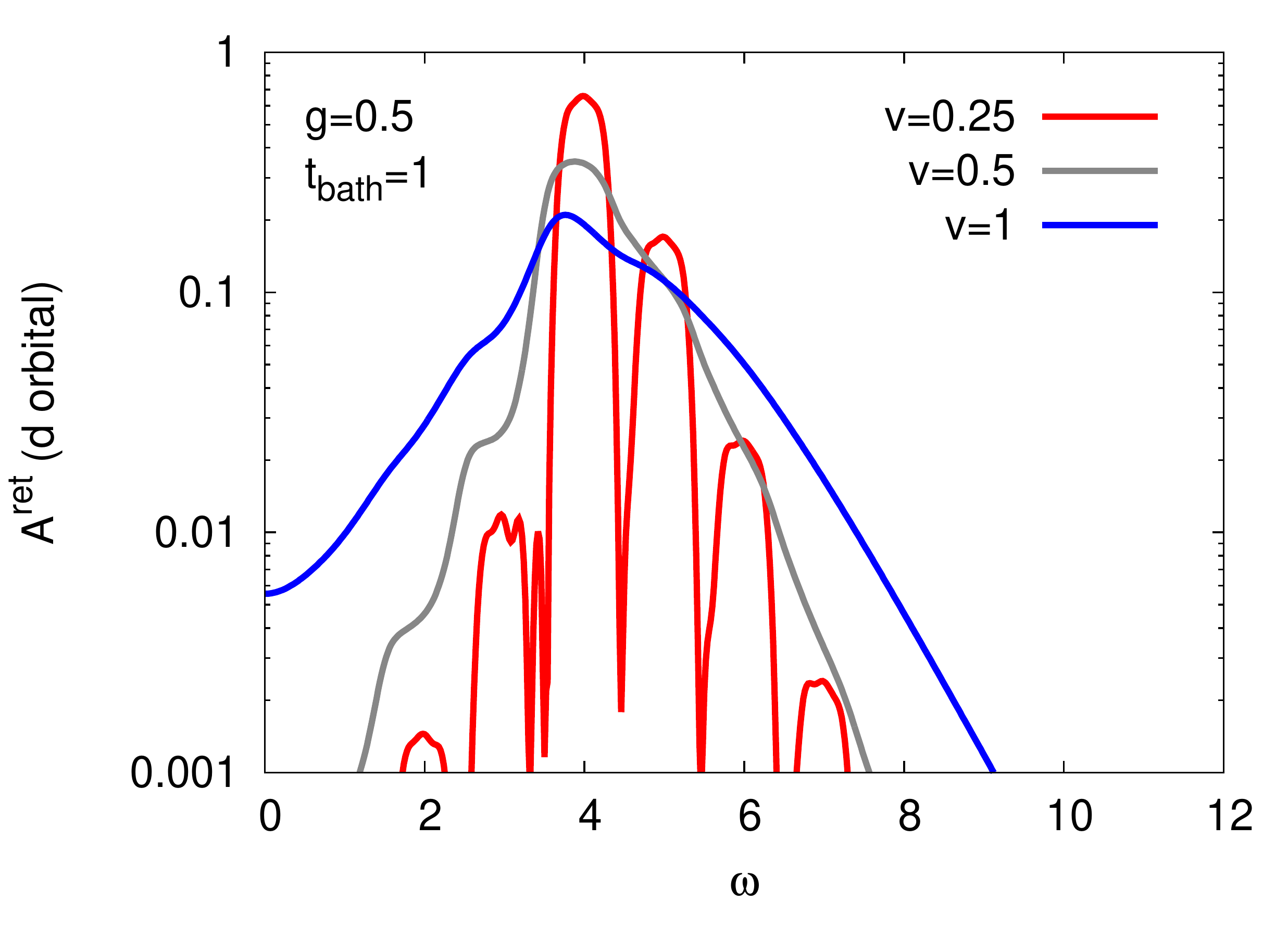} 
\caption{RIXS spectrum at $\omega_\text{in}-|E_\text{core}|=4$ and spectral function (right panels) of the system with $g=1$ (top panels) and $g=0.5$ (bottom panels) after a photo-doping pulse with frequency $\Omega_\text{pump}=8$, for indicated values of the $d$-electron hopping $v$ and core-bath coupling $t_\text{bath}$. }
\label{fig_pump}
\end{center}
\end{figure*}

\subsection{Nonequilibrium results}
\label{sec:results:noneq}

We will consider two different protocols to drive the Mott insulating system out of equilibrium: (i) the application of a pulsed electric field, and (ii) the application of a static electric field. The first set-up describes a photo-doped system if the frequency of the pump pulse satisfies $\Omega_\text{pump}\approx U_{\text{scr}}$. The second set-up mimics the effect of a strong THz field pulse with a frequency below the gap, and much smaller than the inverse hopping time.

\paragraph{Pulsed field.} 
We consider a photo-doped system, in which doublon-hole pairs are created by an electric field pulse  applied to the $d$-subsystem before the RIXS measurement. The pump pulse has frequency $\Omega_\text{pump}=U_{\text{scr}}=8$, a maximum amplitude $E_\text{pump}=8$ (unless otherwise noted) and a Gaussian envelope $f_\text{pump}(t-t_\text{pump})=\exp(0.7(t-t_\text{pump})^2)$, so that the pulse contains about seven cycles. It is centered at $t_\text{pump}=4$, while the RIXS probe pulse is centered at $t_\text{probe}=8$.  

Since doublons can hop between sites and thereby emit (and also absorb) phonons, the photo-doped system has an excited phonon population. As seen in the top right panel of Fig.~\ref{fig_pump}, which plots the nonequilibrium spectral function, this leads to additional phonon sidebands in the gap region. These sidebands are populated by doublons ($\omega>0$) and holons ($\omega<0$) and are qualitatively similar to the sidebands observed in chemically doped Mott insulators.\cite{Werner2015} For the smaller phonon coupling (lower right panel), individual sidebands cannot be resolved for the smaller two values of $g/v$, but one still observes a partial filling of the gap in the photo-doped state. 

The left three panels of the figure show the corresponding RIXS spectra for $\omega_\text{in}-|E_\text{core}|=4$. 
There is no enhancement of the phonon sidebands, compared to the equilibrium spectrum, since the pump field is off during the RIXS measurement, and hence there is no band renormalization as one might expect in a periodically driven Floquet state.\cite{Dunlap1986,Holthaus1992,Tsuji2011} However, the photo-doping and the excited phonon population leave some traces. The density of doublons $D=\langle n_{d\uparrow}n_{d\downarrow}\rangle$ produced by the photo-doping depends strongly on the coupling $g$ and the hopping $v$, being particularly large for large $v$ and small $g$, see Tab.~\ref{tab_D}. A large $D$ leads to a bleaching effect, since doublons already residing on the $d$ orbital block excitations from the core level. On the other hand, 
these doublons contribute to phonon sidebands of the resonant peak and $d$-$d$ excitation peak, because processes where doublons (and holons) hop in or out between the RIXS photon absorption and emission are enhanced. These hoppings can emit phonons and result in an emission with energy reduced by multiples of the phonon frequency. 
We have measured the dependence of the second sideband of the elastic line ($\omega_\text{out}-|E_\text{core}|=2$) on the photo-doping concentration for $v=0.5$, $t_\text{bath}=0.5$ and found for both values of $g$ that the corresponding RIXS signal scales linearly with $D$, at least up to $D\approx 0.08$.

Holes in the $d$ orbitals furthermore activate new processes near the incoming photon energy $\omega_\text{in}-|E_\text{core}|=-4$, as is seen by comparing the top and middle panels of Fig.~\ref{fig_rixs_3d}. Both an elastic feature near $\omega_\text{out}-|E_\text{core}|=-4$, associated with the transient creation of singlons, and a gain feature near $\omega_\text{out}-|E_\text{core}|=4$ associated with doublon-holon recombination processes can be found. In the latter process a holon is converted into a singly occupied $d$ site by the core-valence excitation. This singlon is subsequently converted into a doublon by an electron hopping in from a neighboring doublon, while the de-excitation leaves behind a singlon: $(\uparrow\downarrow , 0) (\uparrow\downarrow , \uparrow\downarrow)\xrightarrow{\omega_\text{in}}$ $(\uparrow , \downarrow) (\uparrow\downarrow , \uparrow\downarrow)$ $ \xrightarrow{\text{hop}}(\uparrow , \uparrow\downarrow) (\uparrow\downarrow,\downarrow)   \xrightarrow{\omega_\text{out}} (\uparrow\downarrow , \uparrow) (\uparrow\downarrow , \downarrow)$. Such processes result in a signal at $\omega_\text{out}-\omega_\text{in}\approx 8$, due to the energy gain of $U_\text{scr}=8$ from the doublon-holon recombination. 

\begin{table}
\begin{tabular}{l|llll}
& $g=1$ & &$g=0.5$ & \mbox{}  \\
\hline
$v=0.25$\hspace{5mm} & 0.0090 \hspace{2mm}\mbox{} & (0.00033) \hspace{5mm}\mbox{}& 0.027 \hspace{2mm} & (0.00044) \\
$v=0.5$ & 0.033 & (0.0013) & 0.085 & (0.0018) \\
$v=1$ & 0.058 & (0.0052) & 0.16 & (0.0070) \\
\end{tabular}
\caption{Double occupation $D$ after the photo-doping pulse. The initial equilibrium values are shown in brackets. Since these results are very weakly dependent on $t_\text{bath}$, we only show the results for $t_\text{bath}=1$.}
\label{tab_D}
\end{table}

\begin{figure*}[ht]
\begin{center}
\includegraphics[angle=0, width=0.49\columnwidth]{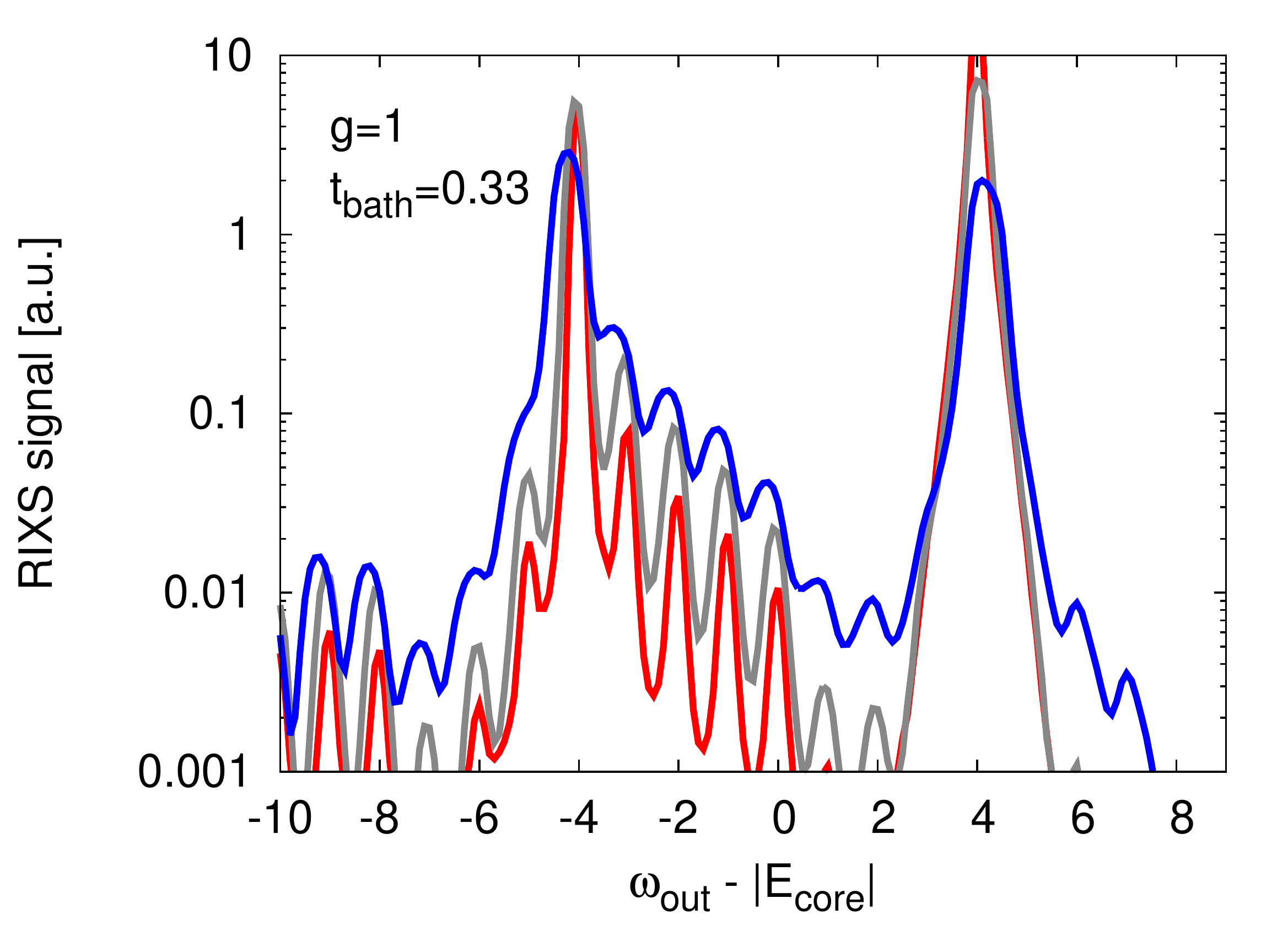}\hfill 
\includegraphics[angle=0, width=0.49\columnwidth]{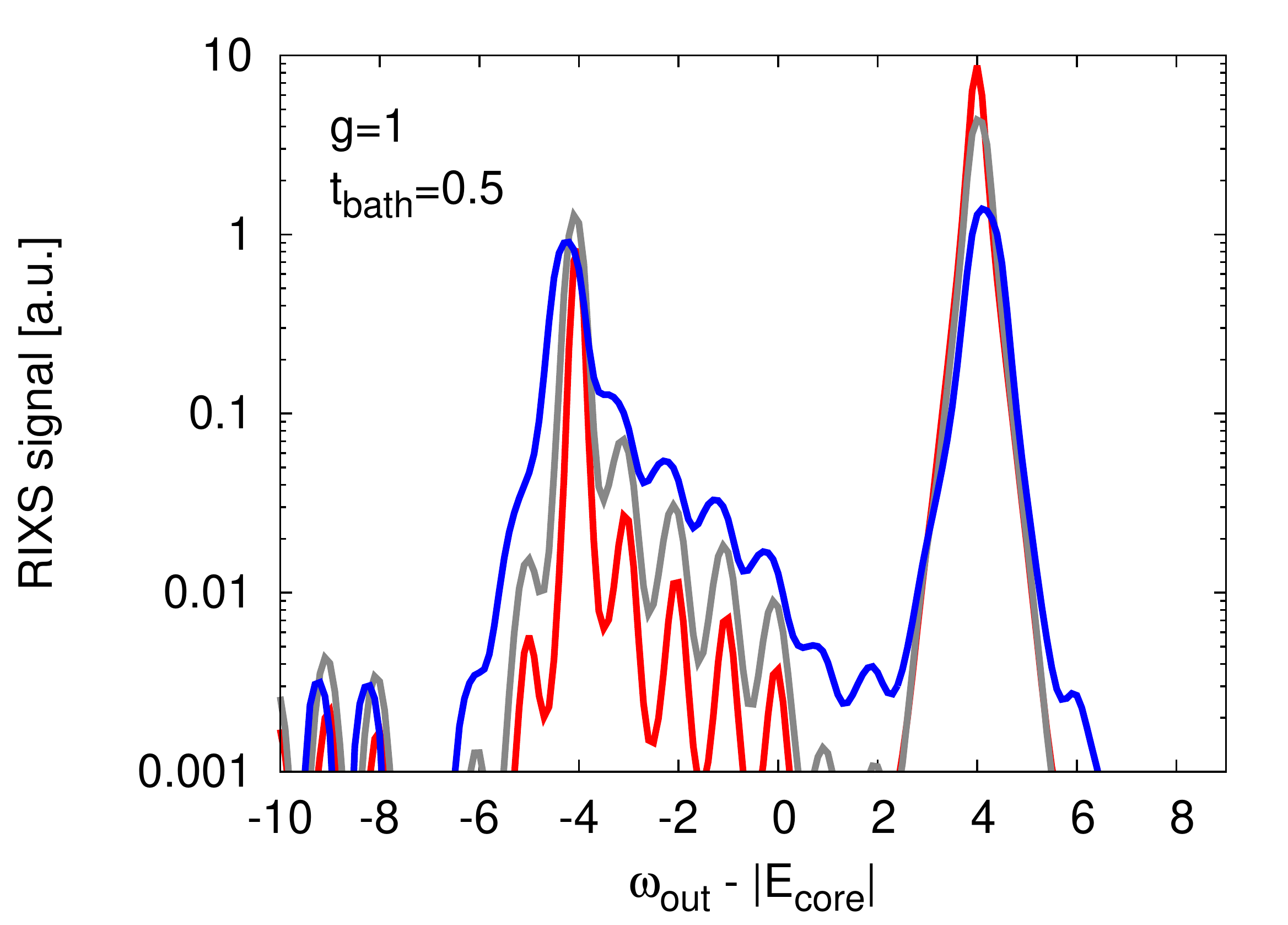}\hfill 
\includegraphics[angle=0, width=0.49\columnwidth]{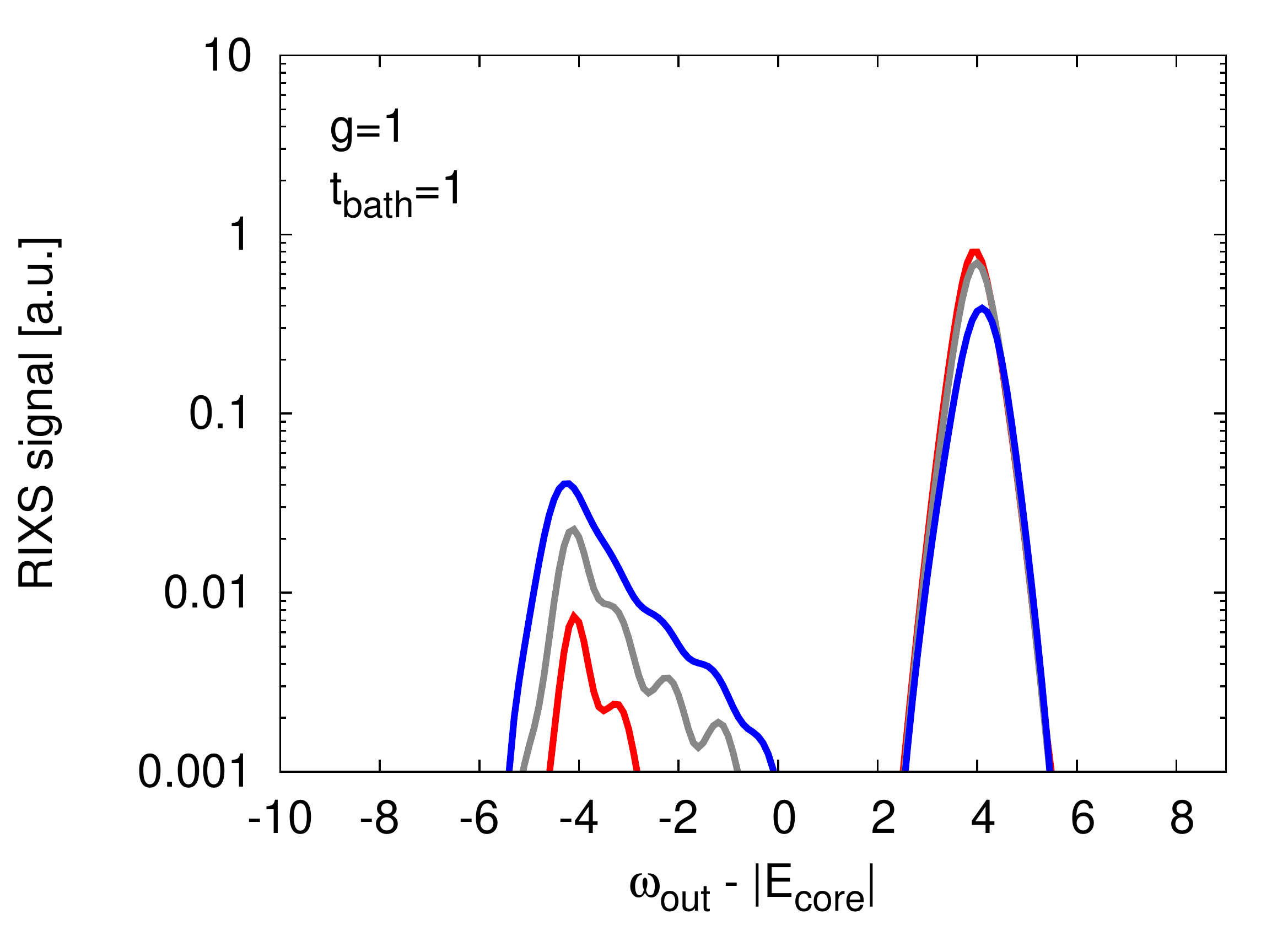}\hfill 
\includegraphics[angle=0, width=0.49\columnwidth]{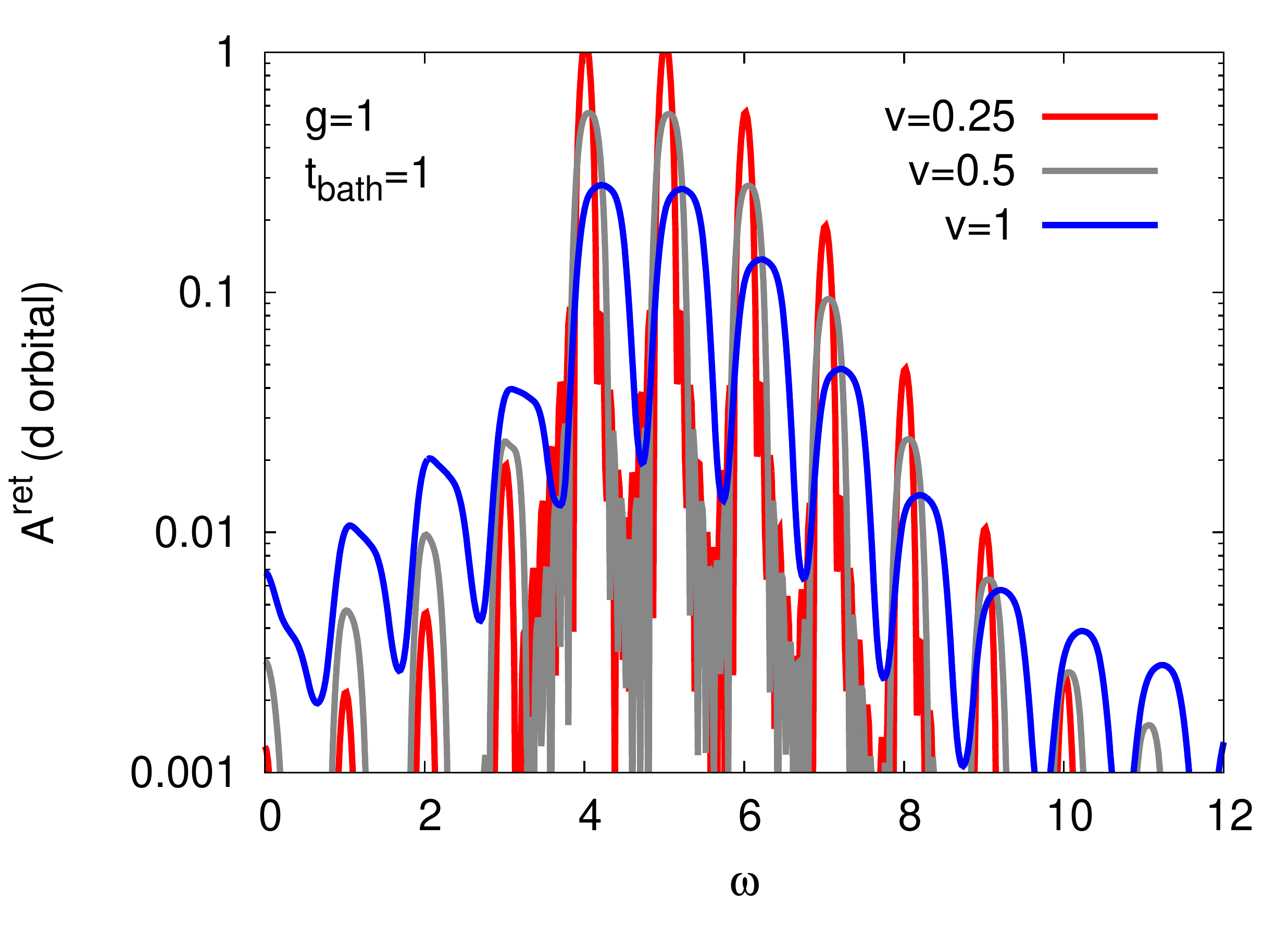}\\ 
\includegraphics[angle=0, width=0.49\columnwidth]{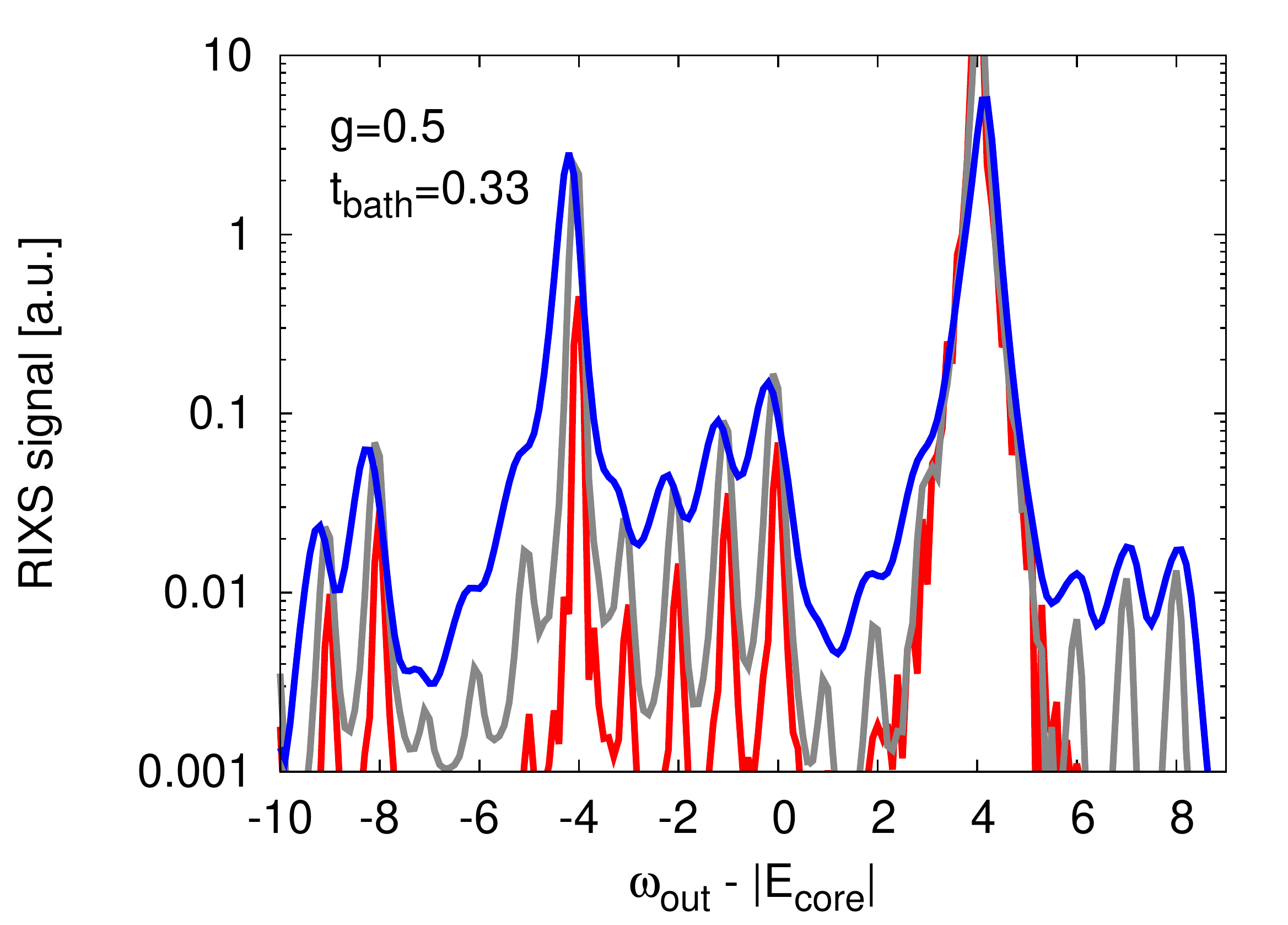}\hfill 
\includegraphics[angle=0, width=0.49\columnwidth]{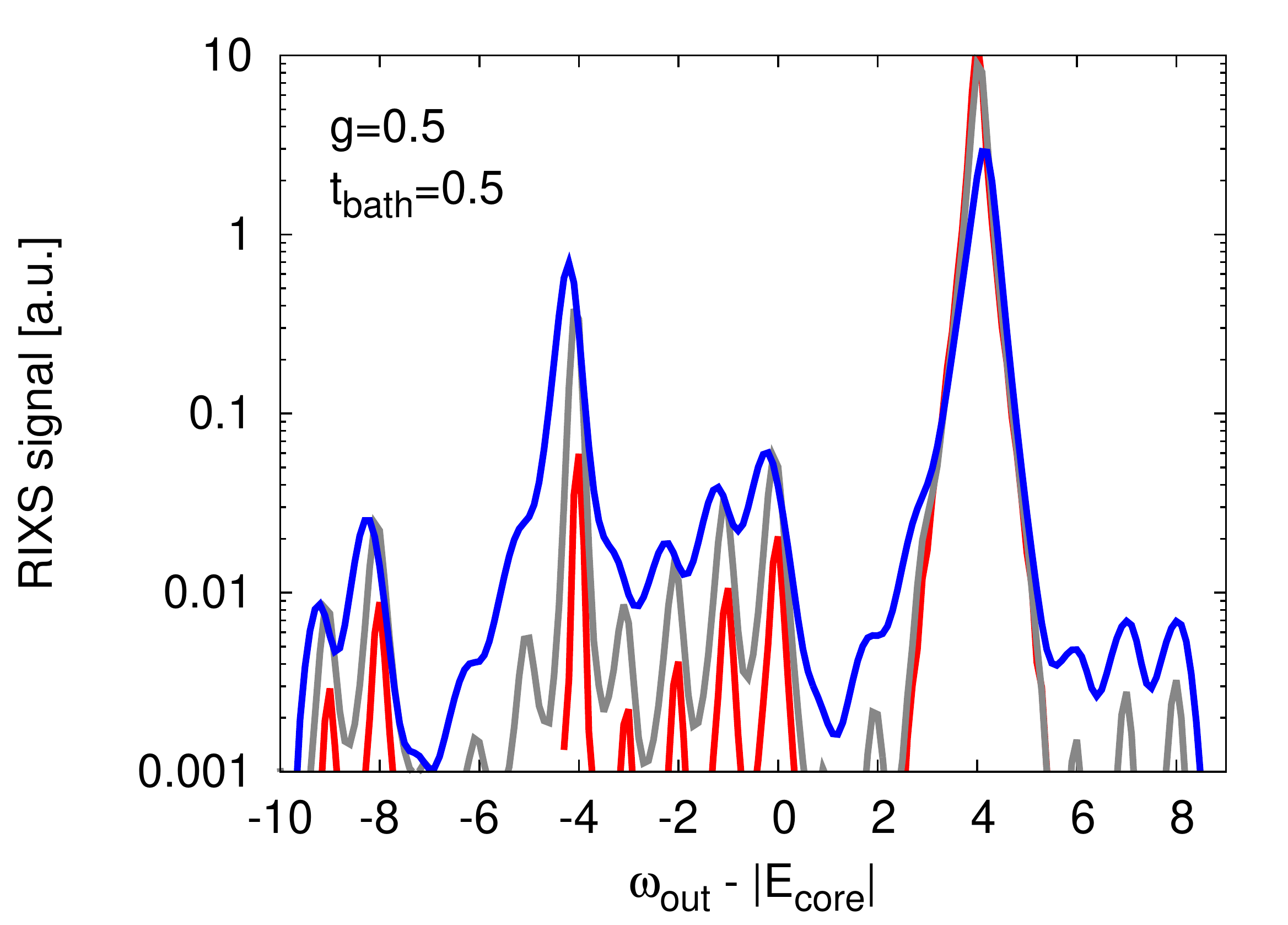}\hfill 
\includegraphics[angle=0, width=0.49\columnwidth]{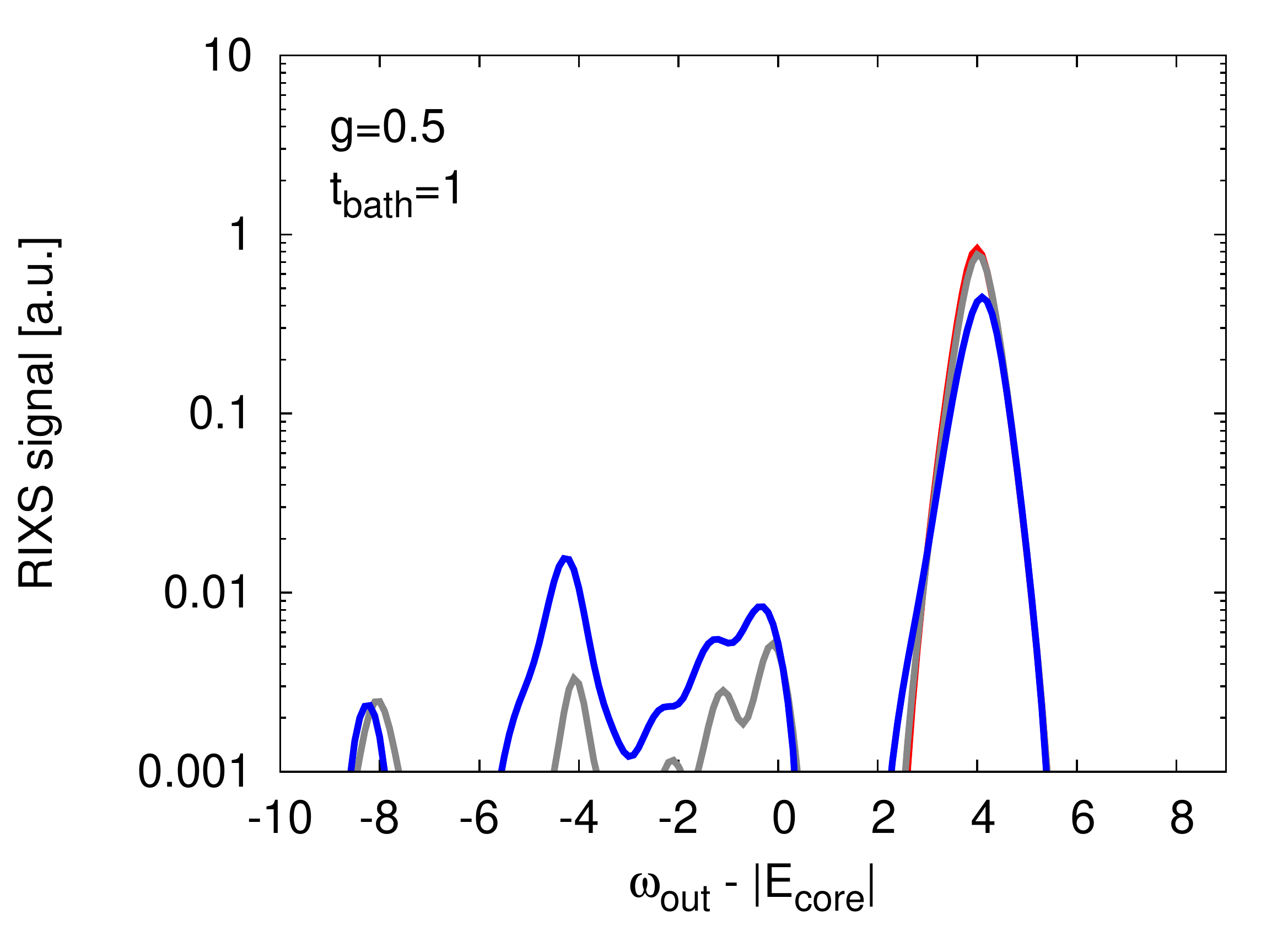}\hfill 
\includegraphics[angle=0, width=0.49\columnwidth]{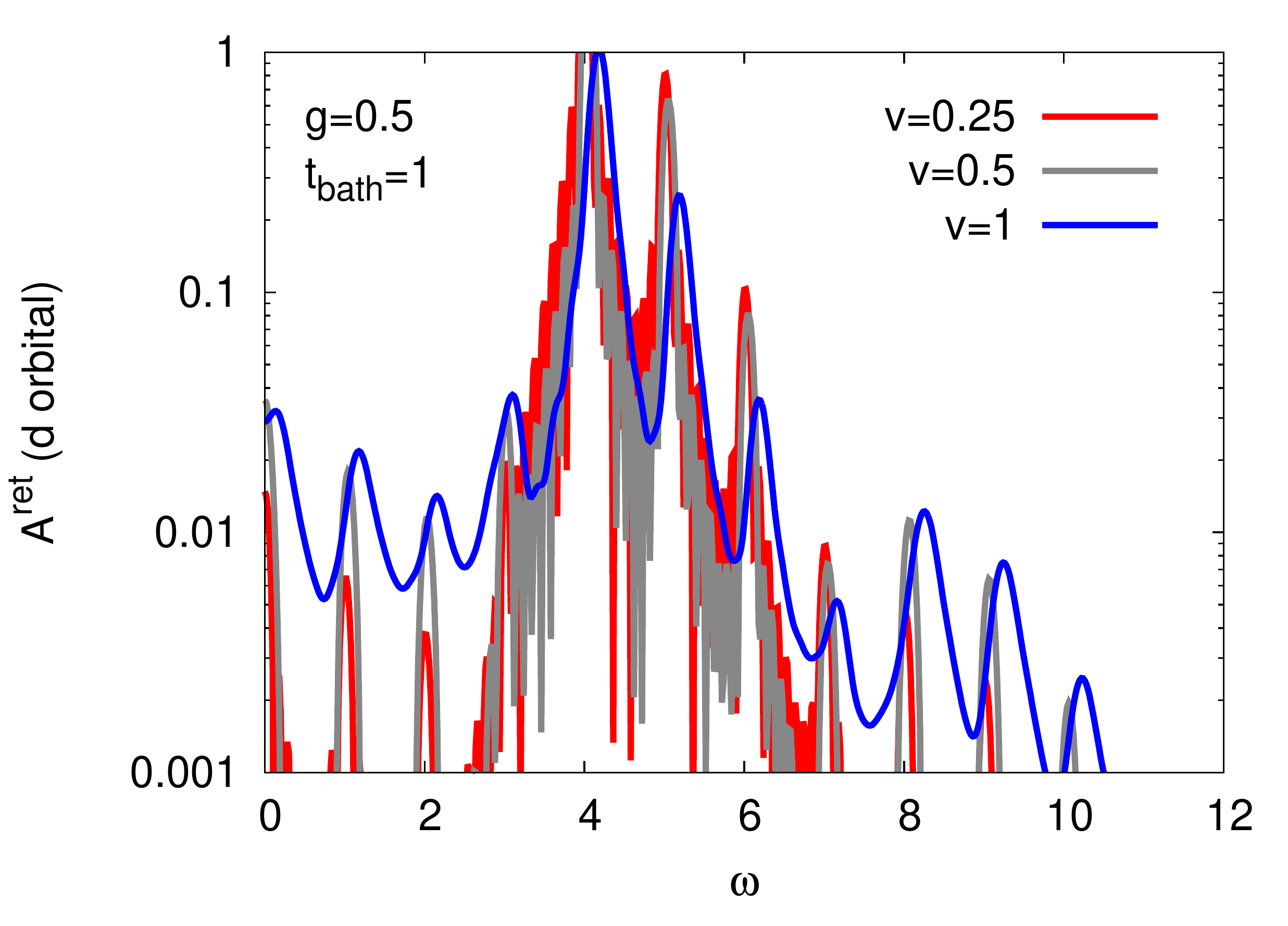} 
\caption{RIXS spectrum at $\omega_\text{in}-|E_\text{core}|=4$ and spectral function (right panels) of the system with $g=1$ (top panels) and $g=0.5$ (bottom panels) under an applied static electric field $E=4$, for indicated values of the $d$-electron hopping $v$ and core-bath coupling $t_\text{bath}$.}
\label{fig_static}
\end{center}
\end{figure*}

\paragraph{Static field.} 
Next, we smoothly switch on a static electric field $E=4$ within a time $\tau=4.5$ and measure the spectral function and RIXS signal shortly after this ramp-up ($t_\text{probe}= 8$). The results for different $v$ and $t_\text{bath}$ are shown in Fig.~\ref{fig_static}. While a static electric field also produces a field-induced tunneling across the Mott gap,\cite{Oka2005,Eckstein2010} and hence an enhanced doublon population, this effect does not play the dominant role here. For $g=1$, the doublon density in the middle of our simulated time interval (where a large fraction of the RIXS signal has already been emitted) reaches $D(t=22.5)=0.0012$, $0.0070$, $0.0446$ for $v=0.25$, $0.5$, $1$, respectively. The corresponding doublon densities for $g=0.5$ are $0.0089$, $0.072$, and $0.121$. These values are relatively large because of the resonant condition $E=U_\text{scr}/2$. Nevertheless, as one can see by comparing the results in Fig.~\ref{fig_static} to those in Fig.~\ref{fig_pump}, the spectra are qualitatively different, and one can easily identify features which are specific to the static-field case. 

Focusing first on the spectral function, we notice two main effects. On the one hand, the static field suppresses the hopping due to Wannier-Stark localization,\cite{Freericks2008,Murakami2018} which leads to narrower phonon sidebands for large $g/v$, and the appearance of sharp phonon sidebands even for small $g/v$ (compare the curves for $g=0.5$ and $v=1$ in the right panels of Figs.~\ref{fig_eq} and \ref{fig_static}). The second effect is the appearance of additional prominent sidebands in the gap region and on the high-energy side of the upper Hubbard band. These can be understood as a combination of Wannier-Stark and phonon sidebands.\cite{Werner2015} In particular in the lower-right panel, one can clearly identify two prominent features at energy $U_\text{scr}/2+E=8$ and $U_\text{scr}/2-E=0$, which are Wannier-Stark peaks associated with the insertion of an electron, and the subsequent hopping of the resulting doublon to a neighboring site, either parallel or antiparallel to the static field. For the present parameters ($E=4\omega_0$), these Wannier-Stark features overlap with phonon sidebands, and they themselves spawn phonon-sidebands, associated with processes where the hopping electron emits $n=1,2,\ldots$ phonons.  

The interesting question now is if and how these field-induced features show up in the RIXS spectrum. The results for $\omega_\text{in}-|E_\text{core}|=4$, shown in the left three panels of Fig.~\ref{fig_static}, indicate that the elastic peak is enhanced, relative to the loss feature, by the static field. This effect can be expected if electrons get localized by the field for the duration of the core hole life-time. The difference between the elastic peaks for different $v$ is reduced and they all resemble the equilibrium result for the smallest $v$, consistent with Wannier-Stark localization, while there are only very weak phonon-related features. 

The effect of the static field on the $d$-$d$ excitation feature is more significant. The phonon sidebands at $\omega_\text{out}-|E_\text{core}|\lesssim -4$ are suppressed, but new, prominent phonon sidebands appear in the energy range $-4\lesssim \omega_\text{out}-|E_\text{core}|\lesssim 0$. This indicates that  new charge excitation processes associated with Wannier-Stark sidebands are activated. In fact, the additional sidebands appear up to an energy $\omega_\text{out}-\omega_\text{in}=-U_\text{scr}+E$.  At this energy, we expect a feature related to the Wannier-Stark sideband at $\omega=U_\text{scr}/2-E$ in $A^\text{ret}$.  More specifically, the energy $\omega_\text{out}-|E_\text{core}|=-4+E=0$ corresponds to a RIXS process, where a core electron is excited to the $d$ orbital, and the resulting doublon hops to the neighboring site in the direction of the applied field (energy gain $E$), before the core hole is filled by the electron left behind.  What the spectra in Fig.~\ref{fig_static} show, especially for $g=0.5$, are phonon sidebands associated with these processes: hopping to the neighboring site in the direction of the field plus $n=0,1,2,\ldots $ phonon emissions. The less prominent sidebands of the main $d$-$d$ excitation peak at $-4$ correspond to electron hoppings without energy gain from the field. Within our model, this requires higher-order processes, and hence the phonon sidebands of the $d$-$d$ feature are suppressed compared to the equilibrium case shown in Fig.~\ref{fig_eq}. (Also the sidebands of the elastic line are related to higher-order processes.) Below energy $\omega_\text{out}-|E_\text{core}|=-4-E=-8$ we can recognize an additional weak feature associated with the other Wannier-Stark sideband (hopping of the doublon against the field, see peak at $\omega\approx 8$ in $A^\text{ret}$), and its phonon sidebands. 

Since for the chosen parameters, the phonon frequency is commensurate with the field, the sidebands associated with the different types of hoppings (energy gain, energy loss, no energy change due to $E$) overlap. For generic field strengths, these different contributions can be shifted relative to each other, which may result in more complicated and blurred spectra. Nevertheless, these examples show that even in the weak phonon coupling regime, where phonon features cannot be resolved in the spectral function or RIXS spectrum, the effect of field-induced localization allows to strongly enhance these features, and at least for suitably chosen field strengths, the phonon sidebands, and hence the phonon frequencies, can be accurately resolved. From the decay of these sideband peaks with $n$ it should furthermore be possible to extract the phonon coupling strength.\cite{Ament2011}

\subsection{Interplay between Wannier-Stark and phonon sidebands}
\label{sec:results:ws}

The RIXS spectrum for $E=4$ is plotted as a function of $\omega_\text{in}$ and $\omega_\text{out}$ in the bottom panel of Fig.~\ref{fig_rixs_3d}. The features discussed in the previous subsection appear near $\omega_\text{in}-|E_\text{core}|=4$, where an efficient creation of doublons by the RIXS pulse is possible even without energy absorption from the static field. 
Energy gained from the field can thus be added to the outgoing photon energy, which results in the upward shift (relative to the $d$-$d$ excitation peak) of the Wannier-Stark-related feature and its phonon sidebands ($\omega_\text{out}-|E_\text{core}|=0,-1,-2,-3$). 

\begin{figure}[t]
\begin{center}
\includegraphics[angle=0, width=\columnwidth]{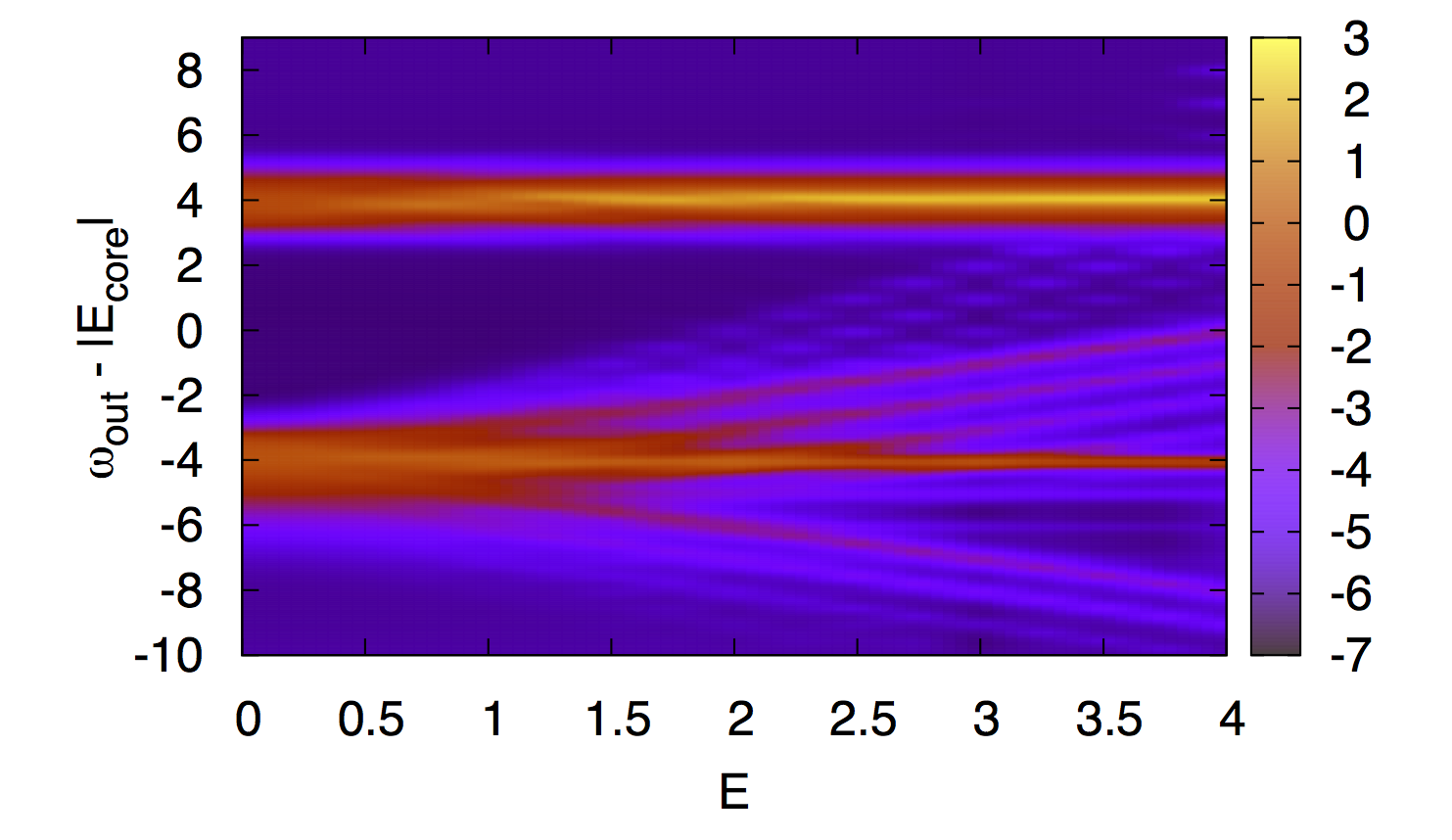} 
\caption{RIXS spectrum for $\omega_\text{in}-|E_\text{core}|=4$ as a function of the applied static field $E$ (log-scale plot in arbitrary units). The model parameters are $g=0.5$, $v=0.5$ and $t_\text{bath}=0.5$.
}
\label{fig_static_3d}
\end{center}
\end{figure}

In addition, one finds in Fig.~\ref{fig_rixs_3d} a broad horizontal loss feature near $\omega_\text{out}-|E_\text{core}|=-4$, and $\omega_\text{in}-|E_\text{core}|\gtrsim 0$, which corresponds to the creation of a doublon-holon pair with the help of the energy $E$ gained by the hopping of the doublon to the neighboring site (in the direction of the field, under the emission of $n=0,1,2,\ldots$ phonons). As in the case of the photo-doped system (middle panel), one furthermore recognizes additional features near $\omega_\text{in}-|E_\text{core}|=-4$. Even though they look weak at first sight, they are actually comparable in intensity to the peaks in the middle panel (notice the different color bars).  These features results from the doublon and holon population produced by the switching-on of the field, and by field-induced tunneling processes,\cite{Eckstein2010} as discussed in Sec.~\ref{sec:results:noneq}$b$.

The RIXS spectra for $\omega_\text{in}-|E_\text{core}|=4$ are plotted as a function of the static field $E$ in Fig.~\ref{fig_static_3d}. This plot makes it obvious that the Wannier-Stark sidebands emerge from the $d$-$d$ excitation peak at $\omega_\text{out}-|E_\text{core}|\approx -4$, and not from the elastic line at $\omega_\text{out}-|E_\text{core}|\approx 4$, which is only affected by the band narrowing effect. Once the field-induced localization is strong enough that the $d$-$d$ excitation peak splits up into phonon sidebands, one clearly recognizes two main Wannier-Stark related features, which shift with energy $\pm E$ relative to the main peak, and their associated phonon sidebands. One furthermore recognizes very weak analogous features shifting with energy $+2E$, which correspond to processes involving two hoppings parallel to the static field (these lines are chopped up into segments because of the limited resolution along the $E$ axis).

In addition to the Wannier-Stark sidebands and their phonon sidebands, there are also weak horizontal features, which are phonon sidebands of the $d$-$d$ peak. As mentioned above, these can be associated with higher-order hopping processes in which  phonons are emitted, but no net energy is gained or lost due to the effects of the static field.

Figure~\ref{fig_static_3d} is reminiscent of the corresponding plot of the nonequilibrium spectral function in Ref.~\onlinecite{Werner2015}, which shows the field-induced narrowing of the Hubbard bands, the emergence of Wannier-Stark sidebands, and their interference with phonon sidebands. The RIXS signal however exhibits a qualitatively different field dependence of the elastic line and the $d$-$d$ excitation peak, which originates from the fact that only the latter feature is associated with electron hopping processes, while the former is associated primarily with local excitation and de-excitation processes which neither couple to the static field, nor (within our model) to the phonons. 

\section{Conclusions} 
\label{sec:conclusions}

We have presented a nonequilibrium RIXS study of a Holstein-Hubbard model in which dispersionless phonons couple to the total charge on a given site. This model can be treated with the nonequilibrium DMFT approach introduced in Refs.~\onlinecite{Eckstein2021} and \onlinecite{Werner2021}, using a hybridization-expansion impurity solver\cite{Werner2007,Werner2013} based on a Lang-Firsov decoupling\cite{Lang1962} of the electrons and phonons. Since the RIXS excitation and de-excitation process does not change the occupation on a given site, it does not directly couple to phonons, and phonon features in the RIXS spectrum appear only via the hopping of electrons between neighboring sites. Within a non-crossing approximation, the effect of the phonons can thus be captured by dressing the hybridization function of the impurity model with an appropriately defined bosonic function.\cite{Werner2013,Werner2015} 

In this model, the phonons manifest themselves in different ways in the spectral function and RIXS spectrum. This becomes most evident in the atomic limit, where the RIXS spectrum has no phonon features at all, while the spectral function splits up into a series of phonon sidebands. In this study, we clarified the properties of the two types of spectra in the lattice case for different hopping parameters, phonon couplings, and core-hole life-times. We considered a large-gap Mott insulator and two types of nonequilibrium set-ups: (i) a photo-doped system with a nonthermal population of doublons and holons, and (ii) a Mott insulator in a static electric field below the dielectric breakdown threshold. 

In the equilibrium case, the RIXS spectrum exhibits two features, an elastic peak without prominent phonon sidebands, and a $d$-$d$ excitation peak associated with the production of a doublon-holon pair. Since the latter process requires the hopping of an electron to a neighboring site, it couples to phonons, which results in a series of phonon sidebands on the low-energy side of the peak (loss of energy to the lattice), at least in the strong electron-phonon coupling regime. In the photo-doped state, also the elastic peak exhibits similar phonon sidebands, since it is now more likely that doublons hop out of and onto the probed site in the time interval between the creation and filling of the core hole. The main qualitative difference between the RIXS spectra of the equilibrium and photo-doped state is however the appearance of two new features at a lower incoming photon energy, which are linked to the presence of holes in the nonequilibrium system: an elastic feature associated with the creation and decay of a singly occupied site, and a gain feature related to doublon-hole recombination. In the latter process, the excited core electron converts an empty $d$ site into a singly occupied site, and an electron with opposite spin hops in from a neighboring doublon before the core hole is filled, leaving behind two singly occupied sites. 

Particularly interesting is the effect of a strong static electric field, whose influence on the electron hopping and spectral function has been previously discussed for the Falikov-Kimball model,\cite{Freericks2008} Hubbard model,\cite{Aaron2012} Holstein-Hubbard model,\cite{Werner2015} and multi-orbital Hubbard model,\cite{Dasari2020} and which has been experimentally explored in GaAs\cite{Schmidt2018} organic conductors.\cite{Ishikawa2014}   
Below the dielectric breakdown threshold, such a field has two main effects on the RIXS spectrum, namely a sharpening of the elastic peak and $d$-$d$ excitation feature for $E\gtrsim \text{bandwidth}$ due to field-induced localization (similar to the effect of reducing the hopping $v$), and the appearance of Wannier-Stark sidebands with an energy splitting of $\pm nE$. The latter are associated with hopping processes over $n$ sites parallel or anti-parallel to the field. Since in the RIXS spectrum, only the $d$-$d$ feature is related to electron hopping,  Wannier-Stark sidebands emerge only from this feature, not from the elastic line. In the presence of phonon coupling, the Wannier-Stark sidebands themselves produce phonon sidebands. The combined effect of band narrowing and phonon coupling is a series of well-resolved phonon features under an applied static field, even in systems where phonon sidebands are not visible in the equilibrium spectrum. The application of strong quasi-static fields (THz field pulses with frequency much smaller than the gap and the hopping) thus provides a means to extract information on the phonons, and in particular the phonon energy, even in systems with weak phonon couplings. From the decay of the phonon sidebands, it should furthermore be possible to determine the phonon coupling strength.\cite{Ament2011} 

In a future work, it would be interesting to implement a nonequilibrium RIXS scheme with explicit treatment of phonons, since this provides more flexibility in the types of models that can be studied. For example, this approach would allow us to consider a coupling of the phonons to only the $d$ orbital (as in Ref.~\onlinecite{Ament2011}), or the treatment of nonlinear electron-phonon coupling terms,\cite{Grandi2021} including the models which have been considered in the context of the light-driven organic Mott insulator ET-F2TCNQ.\cite{Singla2015}      

\acknowledgements
The calculations have been performed on the beo05 cluster at the University of Fribourg using a code based on the Nessi library.\cite{Nessi} PW acknowledges support from ERC Consolidator Grant No. 724103 and SNSF Grant No. 200021\_196966, and ME support from ERC Starting Grant No. 716648.

\end{document}